\documentclass[useAMS,usenatbib]{mn2e}

\usepackage{graphicx,subfig}

\usepackage{amssymb,amsmath}

\newcommand{\degree}{\ensuremath{^\circ}}
\newcommand{\diff}{\textrm{d}}
\title[The massive elliptical galaxy, NGC 4649]{Using NMAGIC to probe
  the dark matter halo and orbital structure of the X-ray bright,
  massive elliptical galaxy, NGC 4649} 

\author[P. Das et al.]
{\parbox{\textwidth}{Payel Das,$^{1}$\thanks{E-mail: \texttt{pdas@mpe.mpg.de}}
Ortwin Gerhard,$^{1}$
Roberto H. Mendez,$^{2}$
Ana M. Teodorescu,$^{2}$
Flavio de Lorenzi$^{1}$}\vspace{0.4cm}\\
\parbox{\textwidth}{$^{1}$MPI f\"ur Extraterrestrische Physik,
  P.O. Box 1603, 85740 Garching, Germany\\
 $^2$ Institute for Astronomy, University of Hawaii, 2680 Woodlawn
  Drive, Honolulu, HI 96822, USA}}

\date{Accepted -. Received -; in original form -}

\begin{document}

\pagerange{\pageref{firstpage}--\pageref{lastpage}} \pubyear{2011}

\maketitle

\label{firstpage}

\begin{abstract}
  We create dynamical models of the massive elliptical galaxy, NGC
  4649, using the N-body made-to-measure code, NMAGIC, and kinematic
  constraints from long-slit and planetary nebula (PN) data. We
  explore a range of potentials based on previous determinations from
  X-ray observations and a dynamical model fitting globular cluster
  (GC) velocities and a stellar density profile. The X-ray mass
  distributions are similar in the central region but have varying
  outer slopes, while the GC mass profile is higher in the central
  region and on the upper end of the range further out. Our models
  cannot differentiate between the potentials in the central region,
  and therefore if non-thermal pressures or multi-phase components are
  present in the hot gas, they must be smaller than previously
  inferred. In the halo, we find that the PN velocities are sensitive
  tracers of the mass, preferring a less massive halo than that
  derived from the GC mass profile, but similar to one of the mass
  distributions derived from X-rays. Our results show that the GCs may
  form a dynamically distinct system, and that the properties of the
  hot gas derived from X-rays in the outer halo have considerable
  uncertainties that need to be better understood. Estimating the mass
  in stars using photometric information and a stellar population
  mass-to-light ratio, we infer a dark matter mass fraction in NGC
  4649 of $\sim 0.39$ at 1$R_e$ (10.5 kpc) and $\sim 0.78$ at
  4$R_e$. We find that the stellar orbits are isotropic to mildly
  radial in the central $\sim 6$ kpc depending on the potential
  assumed. Further out, the orbital structure becomes slightly more
  radial along $R$ and more isotropic along $z$, regardless of the
  potential assumed. In the equatorial plane, azimuthal velocity
  dispersions dominate over meridional velocity dispersions, implying
  that meridional velocity anisotropy is the mechanism for flattening
  the stellar system.
\end{abstract}


\section{INTRODUCTION}

Massive elliptical galaxies are huge conglomerates of stars, dark
matter and hot gas, often residing at the centres of dense
environments. They are believed to evolve through a complex formation
process, manifested in the intricate structure of their
orbits. Understanding the distribution of mass in massive ellipticals
is vital for obtaining constraints on the dark matter content and on
the orbital structure of the stars.

The hot, low-density gas surrounding massive elliptical galaxies
produces X-ray spectra consisting of emission lines and continuous
emission from thermal bremsstrahlung radiation. The spectra can be
modelled to obtain the density and temperature profiles of the gas. If
the gas is relatively undisturbed then we can assume hydrostatic
equilibrium and derive the total mass distribution from the density
and temperature profiles
\citep[e.g.][]{nuls+95,fuk+06,hump+06,chur+08,nag+09,das+10}.

Another common method for obtaining mass profiles of massive
elliptical galaxies is through the creation of dynamical models. They
can be constructed by superposing a library of orbits
\citep[e.g.][]{rix+97,geb+03,thomas+04,cap+06,vdbosch+08} or
distribution functions \citep[e.g.][]{dejong+96,gerhard+98,kron+00},
or by constructing a system of particles
\citep[e.g. NMAGIC,][]{lorenzi+08,lorenzi+09} such that the projection
of the system best reproduces observed surface-brightness and
kinematic profiles. These give the mass distribution and orbital
structure of the galaxies simultaneously.

Obtaining the mass distribution from X-rays is obviously a simpler way
and this can then be used as input in the dynamical models of
galaxies, therefore mitigating the usual mass-anisotropy
degeneracy. This would provide more stringent constraints on the
orbital structure derived from dynamical models. Before doing this
however, one has to understand why there are discrepancies between
mass profiles determined from X-rays and from dynamical models, and
how significant these discrepancies are. \cite{das+10} compared their
X-ray mass determinations for a sample of six nearby massive
elliptical galaxies to the most recent dynamical mass determinations
with inconclusive results. The most common discrepancy between the two
is in the central $\sim 10$ kpc. The X-ray circular velocity curve,
$V_c^2 = GM/r$ is on average about 20\% lower than the dynamical
result in this region. Based on work by \cite{chur+10}, \cite{das+10}
speculated that the most likely sources of this discrepancy are: 1)
Non-thermal contributions to the pressure that are not considered in
the application of hydrostatic equilibrium, 2) multi-phase components
in the gas, 3) a lack of spatially extended observational constraints
in the dynamical models, and 4) mass profiles in the dynamical models
that are not sufficiently general.

In this paper, we would like to look in more detail at NGC 4649, one
of the massive elliptical galaxies in the sample of \cite{das+10},
residing at the centre of a sub-clump in the Virgo cluster. It is the
fourth brightest early-type galaxy in the cluster. Long-slit
kinematics from \cite{bruy+01} and \cite{pink+03} show a high velocity
dispersion of $\sim 400$ km/s in the centre, and a mean velocity of
$\sim 100$ km/s along the major axis. There is also the recent
catalogue of 121 GC line-of-sight (LOS) velocities from
\cite{hwang+08}, building on the catalogue from \cite{brid+06}. They
found significant rotation in the GC system, of order 141 km/s, and an
average velocity dispersion of 234 km/s.

There are several mass distributions determined for NGC 4649 from
X-ray data using ROSAT data \citep{trinch+97}, Chandra data
\citep{brig+97,hump+06,hump+08}, and a combination of Chandra and
XMM-Newton data \citep{nag+09,chur+10,das+10}. The mass profiles are
similar in the central $\sim 12$ kpc except that of \cite{hump+06},
which is a little higher. Further out they all point towards the
existence of a massive dark matter halo, but the precise value of the
circular velocity varies between $\sim$380--500 km/s at 20 kpc
\citep{hwang+08,das+10}.

There are also several dynamical models in the literature combining
the photometric and kinematic data described above. \cite{brid+06}
used axisymmetric Schwarzschild models and \cite{hwang+08} used Jeans
models to fit GC LOS velocities and long-slit kinematics in the X-ray
potential from \cite{hump+06}. They found an isotropic to a modestly
tangential orbital structure. \cite{shen+10} used axisymmetric
Schwarzschild models to fit additionally kinematics from the Hubble
Space Telescope (HST) and carried out an independent mass
analysis. They confirmed the presence of a supermassive black hole in
the centre and a dark matter halo. Their best-fit mass profile is
higher than the mass profiles derived from X-ray observations in the
central $\sim 12$ kpc, and in the outer parts agrees best with the mass
distribution of \cite{das+10}.

In this work, we use the made-to-measure particle-based code NMAGIC
\citep{lorenzi+07} to create dynamical models of NGC 4649. We use
photometric profiles from \cite{korm+09}, long-slit kinematic data
from \cite{pink+03}, and LOS velocities measured for a new catalogue
of 298 PNe described in a companion paper \citep{teod+11}. The PNe
trace the kinematics out to about 416'', similar to the GCs, but with
more than double the number of tracers. There is also strong evidence
that PNe are good tracers of the stellar density and kinematics in
early-type galaxies \citep{coccato+09}.  We explore mass distributions
based on the determinations of \cite{das+10} and \cite{shen+10}, which
only differ in the central $\sim12$ kpc.

With our models we wish to address the following questions:
\begin{enumerate}
\item{How massive is the dark matter halo in NGC 4649?}
\item{What is the orbital structure of the stars in NGC 4649?}
\item{Is the potential derived from X-ray observations accurate enough
    to determine dark matter mass fractions and the orbital structure
    in the halo of massive elliptical galaxies?}
\item{Are the dark matter mass fractions and orbital structure
  in the halo derived from PNe consistent with that derived from GCs?}
\end{enumerate}
We assume that NGC 4649 is at a distance of 16.83 Mpc \citep{tonry+01}
and therefore $1'' = 82$ pc and 1 kpc $ = 12''$. We also assume an
effective radius, $R_e = 10.5$ kpc \citep{korm+09}. Intrinsic radii
are denoted by $r$ and quoted in kpc, and projected radii are denoted
by $R$ and quoted in arcsec. Radii in cylindrical coordinates are also
denoted by $R$.

In Sections \ref{sec:stars} and \ref{sec:vel}, we describe the
photometric and kinematics constraints we use for the projected
distribution function of the stars. In Section \ref{sec:dynmod}, we
describe the dynamical models we create with NMAGIC and discuss the
implications of our results in Section \ref{sec:discuss}. We end with
our conclusions in Section \ref{sec:conc}.

\section{CONSTRAINTS ON THE DISTRIBUTION OF STARS}\label{sec:stars}

\begin{figure}
\centering
\includegraphics[width=8.6cm,angle=-90]{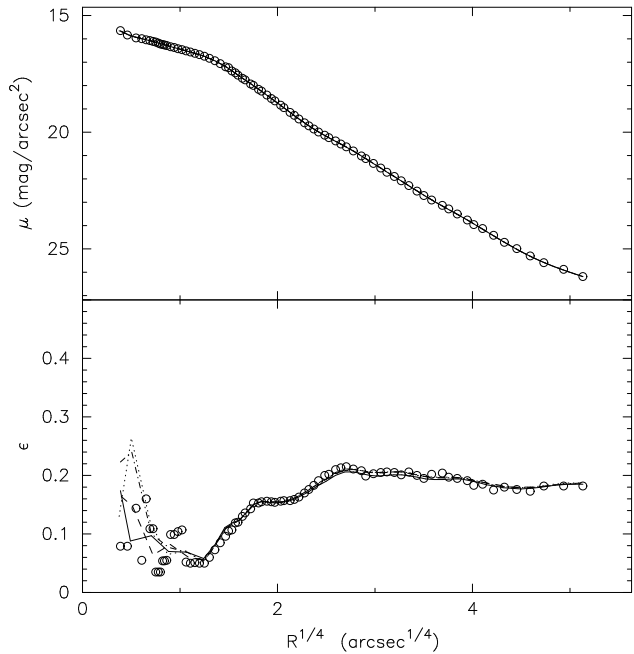}
\caption{Projected light distribution in NGC 4649: The black circles
  in the top panel show the measured $V$-band surface-brightness
  profile and the lines show the reprojected surface-brightness
  profiles from the axisymmetric deprojections. The black circles in
  the bottom panel shows the measured ellipticity profile and the
  lines show the ellipticity profiles of the reprojected light
  distributions from the axisymmetric deprojections. We carry out
  deprojections assuming inclinations of 45$\degree$ (solid),
  60$\degree$ (dashed), 75$\degree$ (dotted), and 90$\degree$
  (dash-dotted).\label{fig:phot}}
\end{figure}

\subsection{Photometric data}\label{sub:photdata}

We use the $V$-band photometry of \cite{korm+09} that combines new
measurements with published profiles, and extends out to 693'' along
the major axis. Figure \ref{fig:phot} shows the measured
surface-brightness and ellipticity profiles. The surface-brightness
profile has a central core with a shallow decay extending to about 4''
and then falls off more steeply outwards. \cite{korm+09} fit a
S\'ersic profile between 5--488'', finding a S\'ersic index,
$n=5.36$. The ellipticity of the isophotes is around 0.1 in the
central 2.5'' but then increases to about 0.2 in the outer parts. The
average position angle (measured from North towards East in the sky)
of the major axis from the profile in \cite{korm+09} is $102\degree
\pm 6\degree$, consistent with the value of 105$\degree$ adopted by
\cite{pink+03} for the kinematic data (see Section
\ref{sub:longslit}). Therefore we also adopt a value of 105$\degree$
for the major axis.

\subsection{3-D density distribution of the stars}\label{sub:deproj}

To obtain the intrinsic distribution of the stars we assume that they
form an oblate, axisymmetric system. We use the code of \cite{mag99}
that finds a smooth axisymmetric density distribution consistent with
the surface-brightness distribution, for some assumed inclination
angle. It chooses the solution that maximises a penalised likelihood,
ensuring a smooth 3-D luminosity density.

We carry out axisymmetric deprojections for inclinations of
$i=45\degree$, $i=60\degree$, $i=75\degree$ and $i=90\degree$. The
minimum inclination angle of $i=45\degree$ is calculated using the
relation between apparent flattening, $q$, intrinsic flattening,
$\xi$, and inclination, $i$\footnote{$i=0\degree$ corresponds to a
  face-on system, and $i=90\degree$ corresponds to an edge-on
  system.}:
\begin{equation}\label{eqn:q}
  q^2 = \cos^2 i + \xi^2 \sin^2 i
\end{equation}
The maximum apparent flattening from the isophotal analysis
is 0.79 and if we assume a maximum intrinsic flattening of 0.5 then
the most face-on inclination allowed is $45\degree$.

Figure \ref{fig:phot} shows the reprojection of the output deprojected
profiles compared with the original input surface-brightness and
ellipticity profiles. They fit the measured profiles very well except
in the central arcsec. Here the ellipticity profile is not so well
defined, because the isophotes are close to circular as a result of
seeing.

\begin{figure}
\centering
\includegraphics[width=8.5cm,angle=-90]{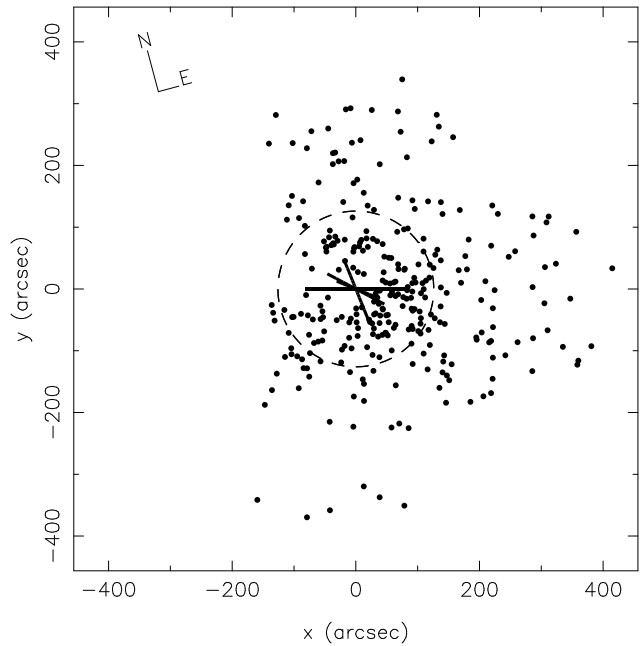}
\caption{Kinematic constraints for NGC 4649: The filled circles show
  the positions of the PNe \citep{teod+11} and the solid lines show
  the positions of the slits. The $x$-axis is along the major axis of
  the galaxy and the $y$-axis is along the minor axis. Directions of
  North and East are shown in the top-left corner. One effective
  radius ($R_e$) is illustrated by the dashed, black
  line.\label{fig:kindata}}
\end{figure}

\begin{figure}
\centering
\subfloat[]{
  \includegraphics[width=1.0\linewidth]{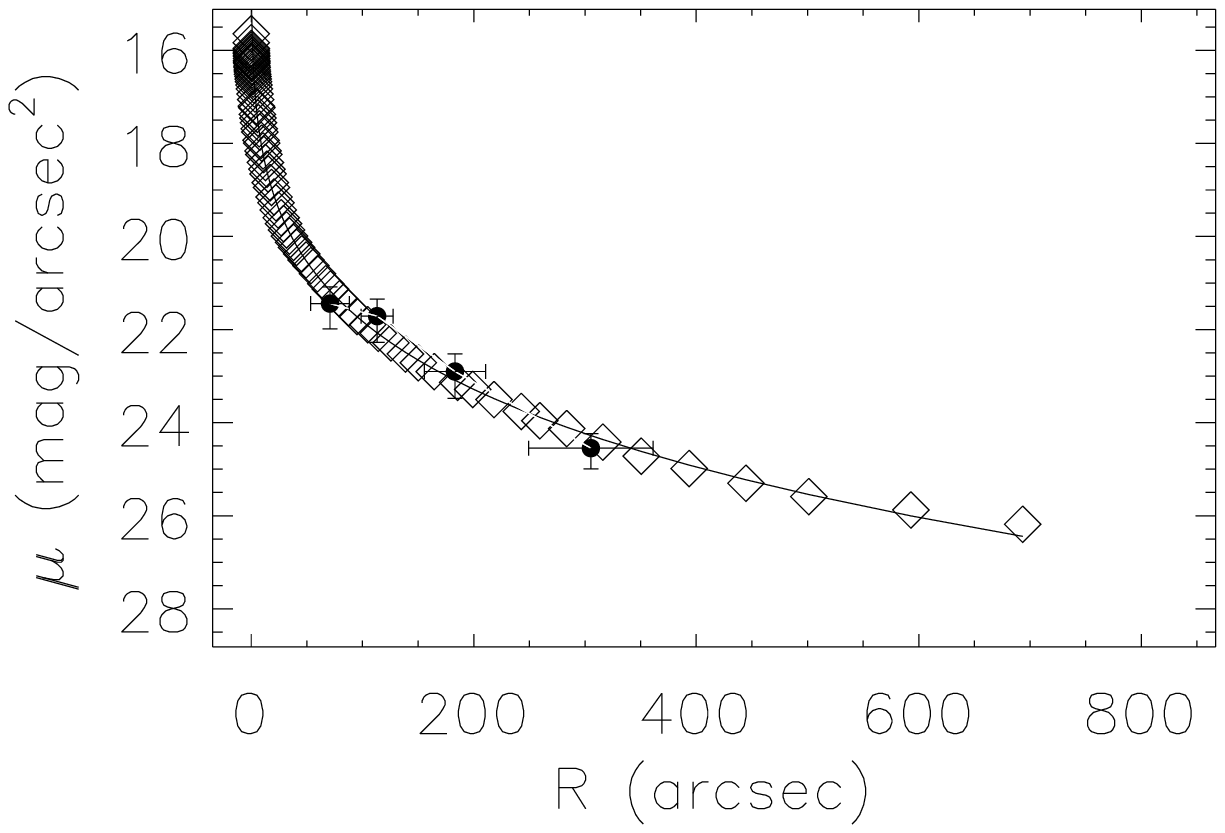}
}\\
\subfloat[]{
  \includegraphics[width=1.0\linewidth]{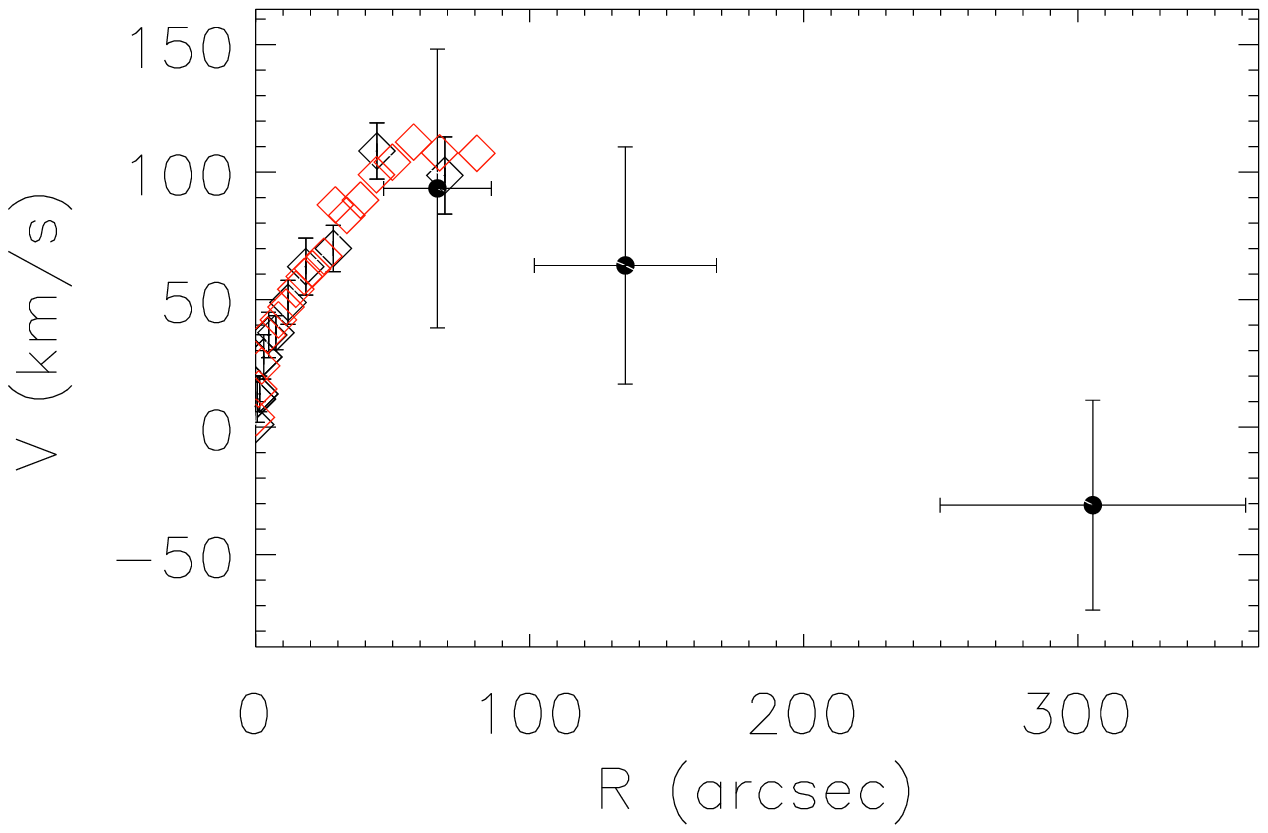}
}\\
\subfloat[]{
  \includegraphics[width=1.0\linewidth]{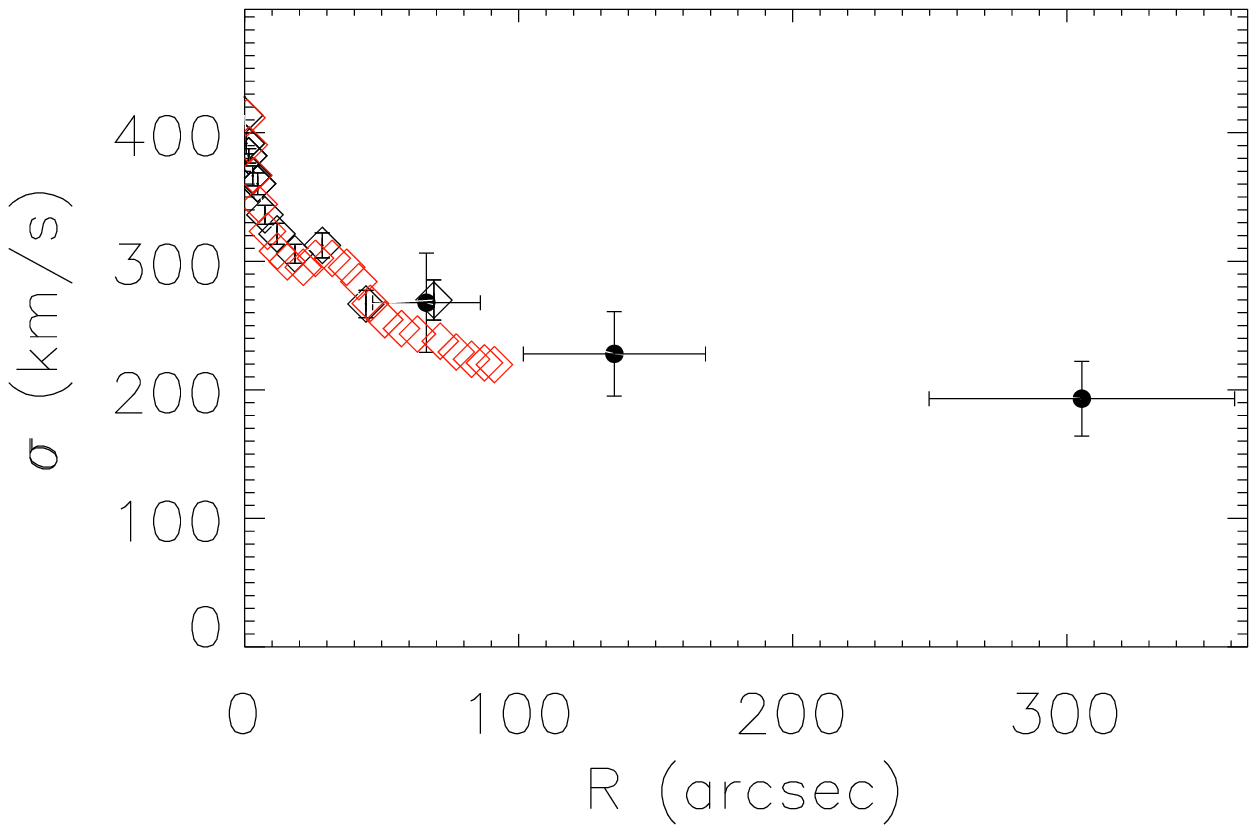}
}
\vspace{0.5cm}
\caption{Comparison of PN data with stellar surface-brightness and
  long-slit data, along the major axis: (a) The surface-brightness
  profile measured from photometry in \protect\cite{korm+09} (black
  diamonds) and the scaled number density calculated from PNe in
  \protect\cite{teod+11} (black circles), along with Poissonian
  errors. (b) and (c) show mean velocity and velocity dispersion
  profiles respectively, measured from long slits in
  \protect\cite{pink+03} (black diamonds), adapted from long-slit
  kinematic data in \protect\cite{bruy+01} (red diamonds, not fitted
  in the NMAGIC models), and from PNe (black
  circles). \label{fig:pne_comp}}
\end{figure}

\section{CONSTRAINTS ON THE DISTRIBUTION OF STELLAR
  VELOCITIES}\label{sec:vel}

As kinematic constraints, we combine long-slit kinematics probing the
central region and planetary nebula (PN) LOS velocities that probe the
outer regions of NGC 4649. Figure \ref{fig:kindata} illustrates the
spatial coverage of the kinematic data. We assume that the $x$-axis
increases along East and that the $y$-axis increases along North, and
then we rotate the coordinate system to align the $x$-axis with the
major axis and the $y$-axis with the minor axis.

\subsection{Long-slit kinematics}\label{sub:longslit}

We use the long-slit absorption-line kinematics of \cite{pink+03},
obtained from STIS/HST measurements and ground-based spectroscopic
measurements on the 2.4m MDM telescope. They were measured along
position angles of 105$\degree$ (the major axis), 127$\degree$,
133$\degree$ and 173$\degree$ measured from North towards East in the
sky, and extending to 69'', 28'', 44'', and 44'' respectively. The
location and orientation of these slits are shown in Figure
\ref{fig:kindata}. \cite{pink+03} derive profiles of the line-of-sight
(LOS) mean velocity, velocity dispersion and the higher-order
Gauss-Hermite moments $h_3$ and $h_4$. Figures \ref{fig:pne_comp}(b)
and (c) show the mean velocity and velocity dispersion profiles
measured by the long-slit data along the major axis. The mean velocity
increases from zero at the centre to about 110 km/s at $\sim 45''$ and
then decreases to about 95 km/s at $\sim 70''$. The velocity
dispersion decreases from about 410 km/s at the centre to about 270
km/s at $\sim 70''$. For comparison we have overplotted major-axis
long-slit kinematic data from \cite{bruy+01}, averaged over both sides
of the galaxy. The agreement is very good. The data from
\cite{bruy+01} are not used in the NMAGIC modelling.

\begin{figure*}
\centering
\subfloat[]{
  \includegraphics[width=0.5\linewidth]{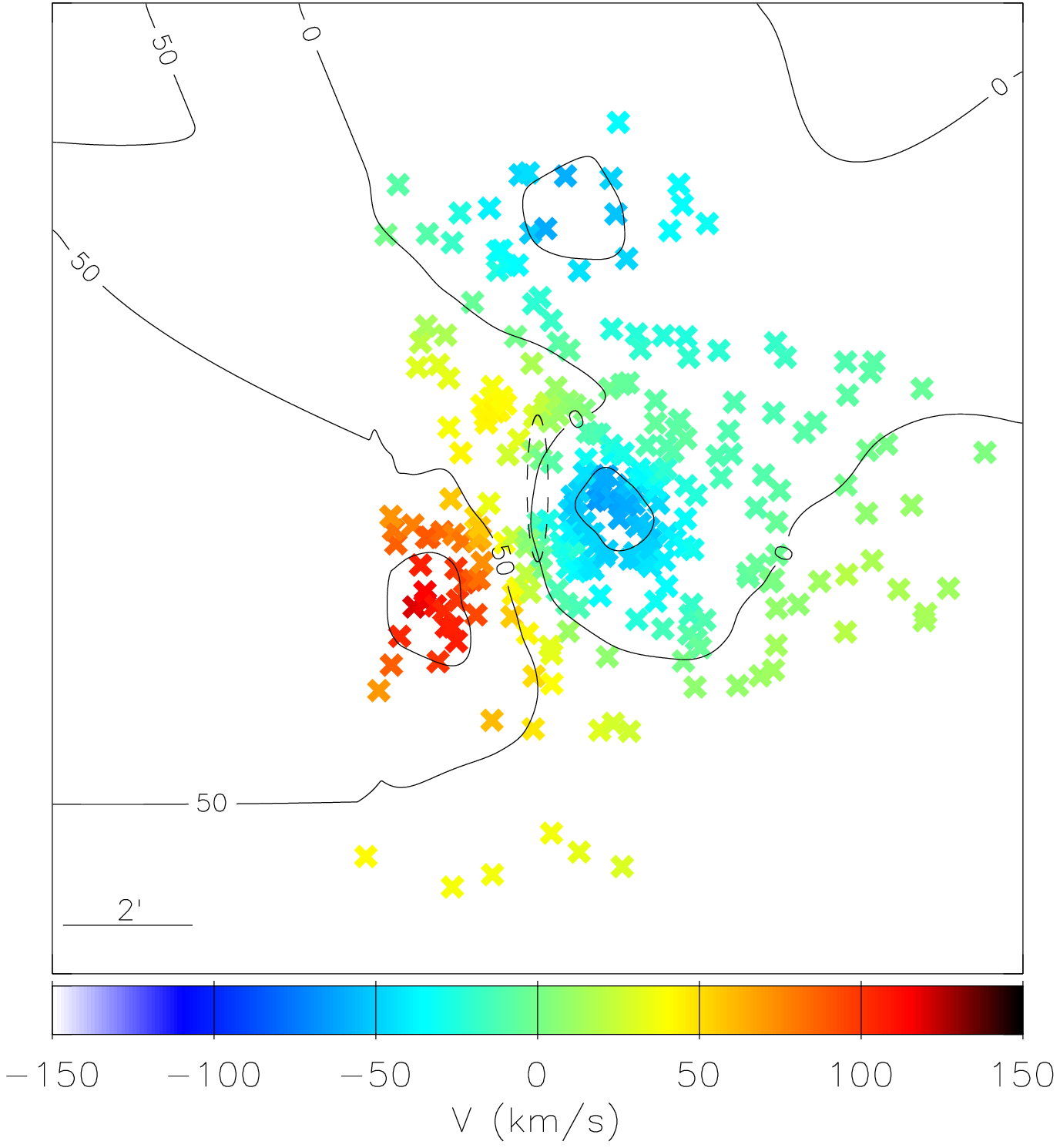}
}
\subfloat[]{
  \includegraphics[width=0.5\linewidth]{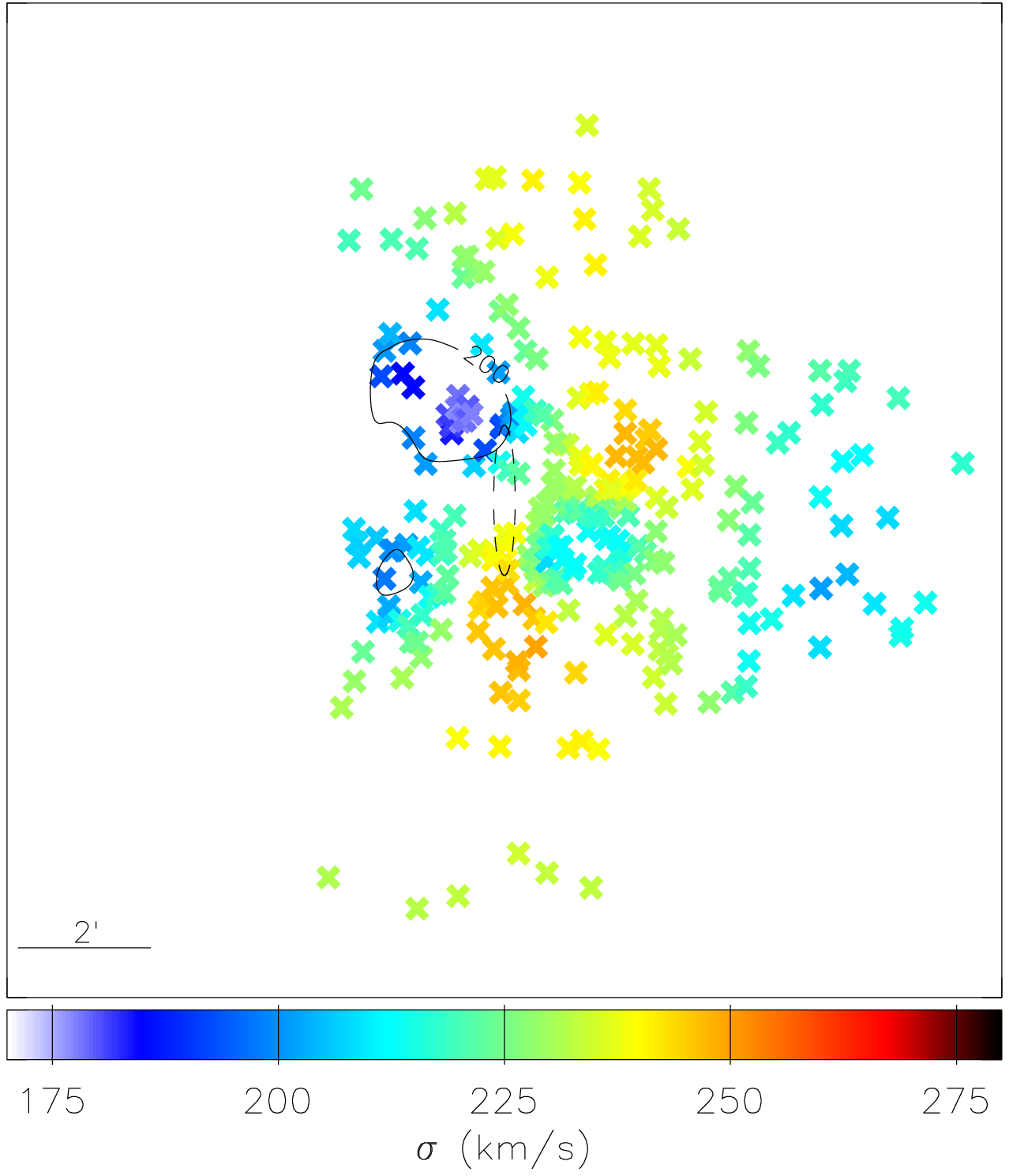}
}
\caption{Smoothed PN kinematic maps: (a) Mean velocity (b) velocity
  dispersion. The maps are orientated as Figure 2, and therefore the
  major axis extends from left to right. \label{fig:smooth_kin}}
\end{figure*}

\subsection{Planetary nebulae}\label{sub:pne}

To enable us to probe the mass and orbital structure in the halo of
NGC 4649, we use LOS velocities derived from observations of PNe with
FORS2 on the VLT and FOCAS on Subaru. The observations and
calculations of the velocities are described in a companion paper
\citep{teod+11}. Contaminants from the neighbouring spiral, NGC 4647,
are removed using the technique developed by \cite{mcneil+10}.  This
results in a catalogue of 298 PNe extending from 27'' to 416'',
therefore overlapping with the long-slit kinematic data.

The location of the PNe are shown in Figure \ref{fig:kindata} and
Figure \ref{fig:pne_comp}(a) compares the measured major-axis $V$-band
surface brightness profile with the scaled number density profile of
the PNe. The PN number density profile is calculated in a cone of
angular width $30\degree$ centred on the major axis, and then scaled
to match the photometry. We do not consider incompleteness
corrections, which are especially important in the central region,
where the bright stellar component masks PNe
\citep[e.g.][]{coccato+09}. However, the PNe number density and
surface-brightness profiles agree well, even at the innermost point.

Figure \ref{fig:pne_comp}(b) shows the mean velocity profile of the
PNe calculated in a 30$\degree$-cone centred on the major axis. In the
region of overlap with the long-slit kinematics, the mean velocity
profiles agree and further out the PN mean velocity profile decreases
outwards to about -20 km/s at $\sim 310''$. The large errors bars
associated with the PN mean velocity profile allow for a much
shallower decrease in the outer parts, consistent with zero at
310''. The combined long-slit and PN major-axis profile shows that the
rotation of the stars increases to about 100 km/s at 50'' and then
starts decreasing outwards.

Figure \ref{fig:pne_comp}(c) shows the velocity dispersion profile of
the PNe calculated in a 30$\degree$-cone centred on the major
axis. The profile agrees with the long-slit kinematic profiles in the
region of overlap and then continues to fall less steeply to a value
of about 200 km/s at $\sim 310$''. The consistent stellar and PN
density and kinematic profiles imply that PNe are good tracers of the
stars both in density and kinematics.

\begin{figure}
\centering
\includegraphics[width=1.0\linewidth]{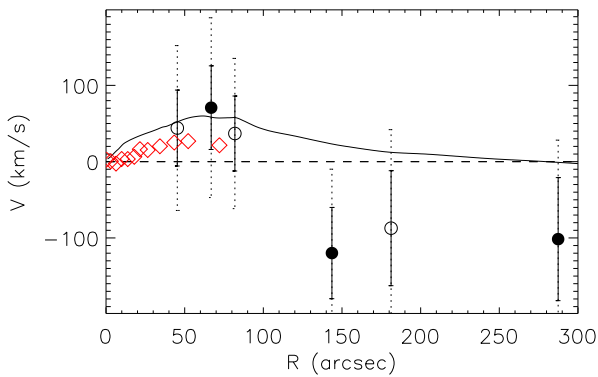}
\caption{Minor-axis mean velocity profile: The filled, black circles
  show the mean velocity profile derived from PNe along a
  30$\degree$-cone centred on the minor axis. The open, black circles
  show the mean velocity profile derived from PNe along a pseudo slit
  of width 60'' centred on the minor axis. The solid black bars denote
  the 1-$\sigma$ errors and the dotted black bars denote the
  2-$\sigma$ errors, estimated from generating pseudo data sets. The
  solid black line shows the mean velocity profile extracted along the
  minor axis of the smoothed mean velocity field derived from the
  PNe. Red diamonds show long-slit kinematic data adapted from
  \protect\cite{bruy+01}. \label{fig:min_vel}}
\end{figure}

Figures \ref{fig:smooth_kin}(a) and (b) show smoothed mean velocity
and velocity dispersion maps for the region of NGC 4649 covered by the
PNe, created by the PNpack suite of IDL routines from
\cite{coccato+09}. Rotation approximately about the minor axis is seen
clearly in (a). There is also a group of PNe in the top of the field
that suggest some rotation about the major axis, a signature of
triaxiality. To investigate the strength of this signature, we first
extract mean velocity profiles for the PNe along a pseudo slit of
width 60'' and a cone of angular width 30$\degree$, centred on the
minor axis. Then to estimate the errors on these profiles, we generate
100 pseudo sets of PN LOS velocities at the same positions as the
original data, assuming that the line-of-sight velocity distribution
(LOSVD) at each point in space is given by the mean velocity and
velocity dispersion of the smoothed kinematic fields. For each of the
pseudo data sets, we extract new mean velocity profiles as done for
the original data set. 1-$\sigma$ and 2-$\sigma$ errors are estimated
from the spread in the mean velocity values at each position. Figure
\ref{fig:min_vel} shows the resulting minor-axis mean velocity
profiles along with the errors. We also extract a mean velocity
profile along the minor axis from the smoothed mean velocity field.
The profile along the pseudo slit is almost consistent with zero
within 1-$\sigma$ errors and both the pseudo-slit and cone profiles
are consistent with zero within 2-$\sigma$ errors. The profile derived
from the smoothed mean-velocity map is very similar within $\sim 90$''
but then decreases to zero, therefore suggesting a lower magnitude of
velocity compared to the other two profiles. This profile should
however be treated with caution in the outer parts as it is based on
far fewer PNe there. The long-slit kinematics of \cite{bruy+01} also
show some rotation along the minor axis at about 20 km/s between
50--80''. We do not plot their last point of $\sim 90$ km/s at the
last radius of $\sim 95''$, as it appears unphysically higher than the
rest of the profile. It would however still be in agreement with the
PN minor-axis mean velocity profiles and is much less than the
measured velocity dispersion. Therefore there is some evidence for
triaxiality but it is weak, making it difficult to constrain the
viewing angles of the system. Therefore we will proceed with models
assuming an oblate, axisymmetric stellar distribution.

\section{DYNAMICAL MODELS}\label{sec:dynmod}

Here we describe how we set up the initial model and how we prepare
the photometric and kinematic target observables for creating
dynamical models with NMAGIC. We then describe how we obtain models
first fitting to the light and long-slit constraints, and then models
fitting light, long-slit and planetary nebula (PN) constraints.

\subsection{NMAGIC}\label{sub:nmagic}

NMAGIC is an N-body made-to-measure code described in
\cite{lorenzi+07} that has been successfully applied to the
intermediate-luminosity elliptical galaxies NGC 4697
\citep{lorenzi+08} and NGC 3379 \citep{lorenzi+09}. It finds the best
intrinsic distribution of stars and their velocities that projects to
fit the photometric and kinematic data. The code builds on the
particle-based made-to-measure method of \cite{syer+96} by accounting
for observational errors and therefore allowing for an assessment of
the quality of the model fits to the target data.  There have also
been recent implementations of this method by \cite{dehnen09} and
\cite{long+10}.

NMAGIC starts with some initial particle model, where each particle
has 3-D spatial coordinates, 3-D velocity coordinates and a
weight. The particles are advanced according to the gravitational
force to sample their orbits and the weights of the particles are
adjusted simultaneously according to the force-of-change (FOC)
equation, given in Equation (10) of \cite{lorenzi+07} and Equation
(22) of \cite{lorenzi+08}. This equation maximises the merit function,
$F$ defined as:
\begin{equation}\label{eq:mf}
F = \mu S - \frac{1}{2}\chi^2 + \mathcal{L}
\end{equation}
where $S$ is a profit function, $\chi^2$ measures the goodness of fit
to the density and long-slit target observables, and $\mathcal{L}$ is
a likelihood term measuring the goodness of fit to the PN target
observables. The parameter $\mu$ changes the balance between the
profit function and the goodness-of-fit terms. For the profit
function, the entropy of the weights is used:
\begin{equation}\label{eq:profit}
S = -\sum_i w_i \ln \left(\frac{w_i}{\hat{w_i}}\right)
\end{equation}
where $\hat{w_i}$ are the initial or prior weights of the particles,
and $w_i$ are the current weights of the particles. The entropy term
causes particle weights to remain close to their priors. Therefore a
higher value of $\mu$ will generally result in smoother intrinsic
distributions. We have chosen a value of $2 \times 10^3$, based on
previous work using NMAGIC \citep{lorenzi+08,lorenzi+09}.

\begin{figure}
\centering
\includegraphics[width=1.0\linewidth]{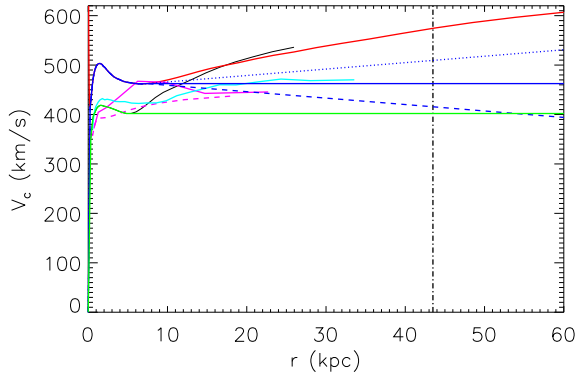}
\caption{Circular velocity curves of NGC 4649: The solid black line
  ($VC1$) shows the best-fit circular velocity curve from
  \protect\cite{das+10}. The solid red line ($VC2$) shows the best-fit
  circular velocity curve from \protect\cite{shen+10}. The blue lines
  show the circular velocity curves based on that from
  \protect\cite{shen+10} but with the outer slope reduced to 0.106
  ($VC3$, dotted), 0.000 ($VC4$, solid) and $-0.106$ ($VC5$,
  dashed). The green line shows the circular velocity curve based on
  that from \protect\cite{das+10} but with the outer slope reduced to
  0.000 ($VC6$). The vertical black dash-dotted line shows the maximum
  radial extent of the kinematic constraints. For comparison, the
  circular velocity curves obtained by \protect\cite{hump+06} (solid,
  pink), \protect\cite{hump+08} (dashed, pink), and
  \protect\cite{nag+09} (cyan) are overplotted. \label{fig:vcirc}}
\end{figure}

\subsubsection{The gravitational force}\label{subsub:force}

In this work we treat the whole gravitational force as an external
force acting on the system, and therefore it can take any form. In the
first instance we consider the circular velocity curves derived for
NGC 4649 by \cite{das+10} ($VC1$) using X-ray observations and
\cite{shen+10} ($VC2$) using optical observations and dynamical
models. These two mass distributions only differ in the central
$\sim12$ kpc and are shown in Figure \ref{fig:vcirc}. The X-ray
observations give information on the temperature and density profiles
of the hot gas in massive elliptical galaxies. If the hot gas is
approximately spherical and in hydrostatic equilibrium, then the
temperature and density profiles are related to the circular velocity
curve by:
\begin{equation}\label{eq:he}
  V_c^2 = - \frac{k_b T}{\mu m_p} \frac{\diff\ln P}{\diff\ln r}
\end{equation}
where $T$ is the temperature of the gas and $r$ is the 3-D
radius from the centre of the galaxy. $\mu = 0.61$ is the average gas
particle mass in terms of the proton mass, $m_p$. This value of $\mu$
corresponds to a helium number density of $7.92\times 10^{-2}$ and 0.5
solar abundance of heavier elements.  We assume that the gas is ideal
and therefore the gas pressure $P = nk_BT$, where $n$ is the particle
number density of the gas. \cite{das+10} applied hydrostatic equilibrium
using a new non-parametric Bayesian approach to density and
temperature profiles of the hot gas derived from Chandra and
XMM-Newton observations \citep{chur+10}.

The best-fit mass profile of \cite{shen+10} was obtained from
axisymmetric Schwarzschild models assuming a stellar density profile
from \cite{korm+09} and fitting to long-slit kinematic constraints
from \cite{pink+03} and GC LOS velocities from \cite{hwang+08}.

In both cases we assume that the gravitational force is spherically
symmetric. Therefore as the stars are assumed to have an oblate,
axisymmetric distribution, the dark matter halo must be prolate in the
centre where the stars dominate, and almost spherical in the outer
parts where the dark matter dominates.

\subsubsection{The initial particle model}\label{subsub:ics_na}

We set up an initial model of 750000 particles extending to $\sim 205$
kpc. We assume a density distribution of the particles given by a
spherical deprojection of the circularly-averaged surface-brightness
profile. We define the intrinsic stellar velocity distribution using
the circularity functions of \cite{gerhard91}, resulting in an
anisotropy profile ($\beta = 1-\sigma_{\theta}^2/\sigma_r^2$) that is
isotropic in the centre but moderately radial ($\beta \sim 0.5$) in
the outer parts. This choice reflects the results of numerical
simulations by \cite{abadi+06} and \cite{onorbe+07}. We then solve for
the energy part of the distribution function assuming a potential
given by $VC1$. The particles' coordinates and velocities are chosen
according to the complete distribution function after \cite{deb+00}
and they are assigned equal weights of $w_i = 1/750000$ to produce the
initial particle model. As the gravitational field is fixed with time
in our models, the orbits of the particles do not change with
time. Therefore our initial particle model is conceptually very
similar to the orbit libraries used as initial conditions in
Schwarzschild models.

\subsubsection{Preparation of stellar density target
observables}\label{sub:photo_na}

We represent the 3-D luminosity density profiles obtained for each
inclination from the deprojection in terms of $A_{lm}$ coefficients
\citep{lorenzi+07} between 0.3 kpc and 201 kpc. 0.3 kpc is the
innermost radius of the X-ray circular velocity curve ($VC1$) and 201
kpc is the radius at which the density of the initial particle model
starts falling off from the deprojected density profile. As the
stellar distribution is assumed to be axisymmetric we set all moments
that describe non-axisymmetry to zero. We calculate the $A_{lm}$s over
a grid of 60 radii. Poissonian errors are assumed for the radial mass
profile and Monte-Carlo simulations are used to determine errors for
the higher-order mass moments.

\begin{table}
  \centering
  \begin{tabular}{c c c c c c}
    \hline
    \hline
    $i$ ($\degree$)   &$\chi^2_{\textrm{alms}}$ &$\chi^2_{\textrm{long-slit}}$  &$\chi^2$        &$-F$ \\
    (1)             &(2)                   &(3)                           &(4)  &(5)\\
    \hline
    45              &0.468,\,0.260         &0.632,\,0.521                
&0.534,\,0.366   &1374.3,1108.0\\
    60              &0.179,\,0.210         &0.537,\,0.469                
&0.324,\,0.315   &746.4,713.0,\\
    75              &0.141,\,0.187         &0.542,\,0.477                
&0.304,\,0.304   &644.6,641.6\\
    90              &0.220,\,0.196         &0.655,\,0.518                
&0.396,\,0.326   &761.5,661.5\\
    \hline
  \end{tabular}
  \caption{Best-fit NMAGIC models for NGC 4649 assuming different
    inclinations in the X-ray ($VC1$, left in columns (2)--(5)) and
    dynamical ($VC2$, right in columns (2)--(5)) potentials : (1)
    Inclination assumed for model, (2) $\chi^2$ per $A_{lm}$ target
    observable, (3) $\chi^2$ per long-slit target observable, (4)
    $\chi^2$ per target observable, and (5) the merit function. There
    are 960 density target observables and 655 long-slit target
    observables. \label{tab:alms+ls_chi2}}
\end{table}

\begin{figure}
\centering
\includegraphics[width=5.0cm,angle=-90]{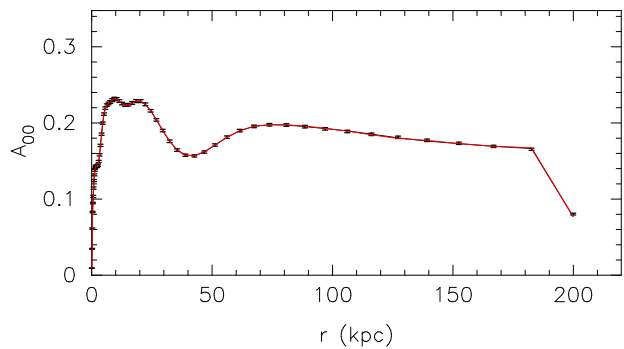}
\caption{Fit to the first moment of the $A_{lm}$s or differential
  stellar mass distribution for an inclination of 75$\degree$ in the
  circular velocity curves $VC1$ (black) and $VC2$
  (red). \label{fig:alms+ls_alms}}
\end{figure}

\subsubsection{Preparation of kinematic target
  observables} \label{sub:kin_na}

As we are assuming that the stellar distribution in NGC 4649 is an
oblate, axisymmetric system rotating about its minor axis, for every
mass element at $(x,y,v)$, we can assume that there is also a mass
element at $(-x,-y,-v)$ and a mass element at $(-x,y,-v)$. Therefore
we can add three more long slits at position angles of
$105\degree-68\degree=37\degree$, $105\degree-28\degree=77\degree$ and
$105\degree-22\degree=83\degree$, which are spatial reflections around
the $x$-axis of the long slits positioned at 173$\degree$,
133$\degree$ and 127$\degree$. The PNe are increased 4-fold by
imposing the above reflections, resulting in a sample of 1192 PNe. As
NMAGIC is a particle-based code, the weight of each particle will be
changed more evenly over each orbit with the 4-fold sample compared to
the unfolded sample.

For NMAGIC the light in each of the slit cells needs to be calculated,
as the light-weighted kinematics are fit rather than the kinematics
directly. The light in the cells is calculated by integrating the
surface-brightness distribution over the dimensions of each of the
slit cells (assumed to have a width of 5'') using a Monte-Carlo
integration scheme.

To fit the PN LOS velocities, we use the likelihood method used in
\cite{lorenzi+08,lorenzi+09}. In this method the particles are binned
in radial and angular segments to calculate the LOSVD in each
segment. Then the likelihood of each PN belonging to the LOSVD of its
segment is maximised for all the PNe.  For the binning, we choose 3
radial bins (corrected for an average projected ellipticity of 0.857)
and 6 equally-spaced angular bins with the first centred on the major
axis.

\begin{figure*}
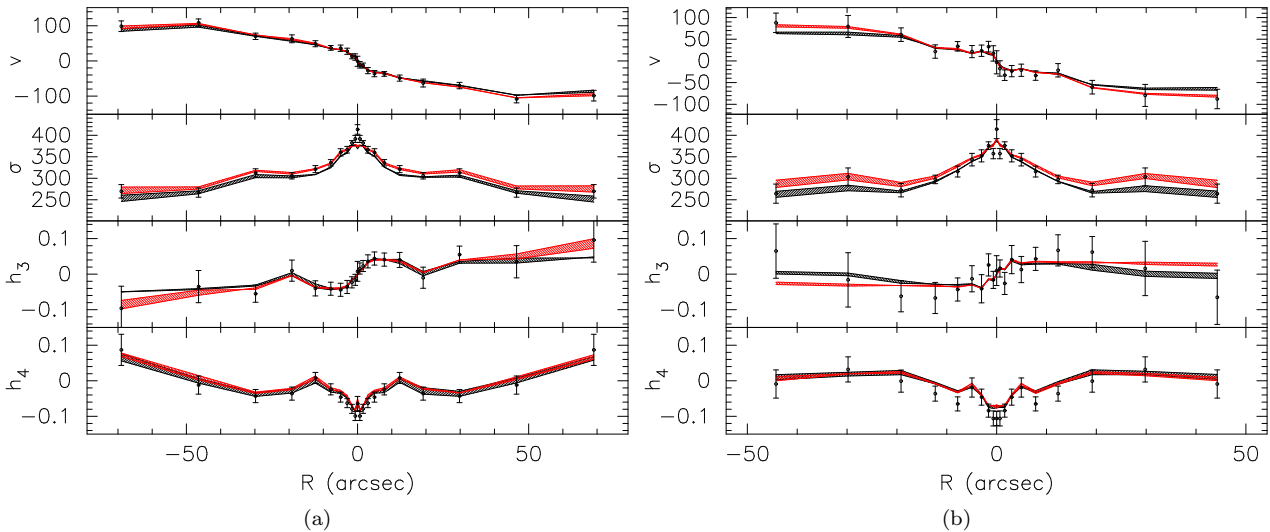

\centering
\subfloat[]{
  \includegraphics[width=6.5cm,angle=-90]{alskin_allinc_alms+ls_maj.ps}
}
\subfloat[]{
  \includegraphics[width=6.5cm,angle=-90]{alskin_allinc_alms+ls_133.ps}
}
\caption{{Fits to the long-slit data (a) along the major axis and (b)
    along a slit placed at a position angle of 133$\degree$: From the
    top down are shown mean velocity, velocity dispersion and the
    Gauss-Hermite moments $h_3$ and $h_4$.} Black circles show the
  data, the black region shows the models for the various inclinations
  probed in $VC1$, and the red region shows the models probed for the
  various inclinations in $VC2$. \label{fig:alms+ls_ls}}
\end{figure*}

\subsubsection{Logistics of a `run'}\label{sub:runs_na}

For each run we fit the desired data starting with the initial
particle model generated above. There is an initial relaxation phase
of 2000 steps where the particles are advanced according to the force
and their weights are not changed. Each step is equivalent to
$4.4\times 10^5$ yrs.\footnote{As we are approaching equilibrium
  systems, the physical time that lapses is less important than the
  number of steps required to achieve it.} We then have another 2000
steps during which we temporally smooth the observables and the
weights of the particles are still not changed. Then the core phase of
the run starts, where the particles are advanced according to the
gravitational force and the weights of the particles are changed
according to the FOC equation. Once the fractional change in the total
$\chi^2$ falls to less than 1.5$\times 10^{-3}$ over 5000 steps, we
define the model to have converged. Finally this is followed by a free
evolution, where the particles are advanced for a further 20000 steps
and the weights are no longer changed, to check that the converged
model has been sufficiently phase-mixed. The particles are advanced
using an adaptive leapfrog scheme.

In the FOC equation, the respective contributions of the $A_{lm}$,
long-slit kinematic and PN kinematic terms can be very different. The
errors are very small for the $A_{lm}$s but much larger for the PNe,
resulting in a much larger contribution from the $A_{lm}$ terms to the
FOC equation. As a result, changes made to the weights of the
particles by the PN data will be small, and need to be made many times
before a converged model is reached. Therefore as the PNe and to some
extent the long-slit are important in the halo, after some tests, we
increase their contributions to the FOC equation by factors of 20 and
2 respectively, to ensure that the halo is adequately modelled.

\subsection{Models fitting density and long-slit kinematic constraints
  only}\label{sub:alms+ls}

First we would like to understand whether the observational
constraints are able to differentiate between the dynamical mass model
of \cite{shen+10} ($VC2$) and the X-ray mass profiles in the
literature, which are lower in the central $\sim 12$ kpc. For the
X-ray mass profile, we choose that of \cite{das+10} ($VC1$), which is
very similar to that of \cite{shen+10} in the outer parts. As the
potentials only differ in the central 12 kpc, and the PN kinematic
constraints are much weaker than the density and long-slit kinematic
constraints, we will not include the PNe for these models. This saves
considerable computational time because the PNe are further out and
therefore the particles need to be integrated for longer there to
achieve convergence. Additionally the likelihood method is itself
computationally expensive.

We carry out NMAGIC models assuming inclinations of 45$\degree$,
60$\degree$, 75$\degree$ and 90$\degree$ for the stellar distribution,
for which we also carried out the deprojections of the
photometry. Table \ref{tab:alms+ls_chi2} shows the $\chi^2$ per data
point for the density observables, the long-slit kinematics, all the
observables, and the merit function $F$ of the final models in each
inclination and potential. NMAGIC fits to the light-weighted kinematic
observables as well as the light in each cell, and therefore the
long-slit $\chi^2$ is calculated as a sum over these variables.

We find that both potentials prefer (in the sense of a lower $\chi^2$)
an inclination of $75 \degree$ of the stellar system. In this
inclination, neither potential is preferred by the combination of the
photometric and long-slit constraints.

\subsubsection{Fits to the observables}\label{subsub:alms+ls_fitobs}

Figure \ref{fig:alms+ls_alms} shows the fits in both potentials to the
first moment of the $A_{lm}$s or differential stellar mass
distribution for the most favoured inclination of 75$\degree$, in
potentials $VC1$ and $VC2$. The fits are almost indistinguishable from
each other, and match the density constraints very well. Figure
\ref{fig:alms+ls_ls} shows the fits to the $v$, $\sigma$, $h_3$ and
$h_4$ moments along the major-axis slit and the slit placed along
133$\degree$, for both potentials and all inclinations. In general the
fits are very similar but there are some systematic differences. The
models in $VC1$ produce mean velocity profiles with a lower magnitude
($\sim 5$--20 km/s), a lower velocity dispersion ($\sim 10$--30 km/s),
an $h_3$ moment with a lower magnitude (0.01--0.02) and an $h_4$
moment that is very marginally lower on average. These differences are
generally a small fraction of the error bars.

All the models find a $\chi^2$ per data point less than 1. Then if the
number of degrees of freedom is approximately equal to the number of
data points, then we can be satisfied that we are fitting the data
well, and even over-fitting. In reality however, the number of degrees
of freedom is difficult to estimate because it is equal to the number
of constraints subtracted by the number of parameters. The number of
parameters is equal to the number of model parameters (e.g. halo,
$M/L$, inclination) plus the number of weights that we are fitting,
which is the number of particles. The number of constraints is equal
to the number of data points plus the number of constraints introduced
by the profit function, which is difficult to quantify.

We now attempt a $\Delta \chi^2$ analysis as done for example in
\cite{shen+10}. If we say each of our models could be characterised by
four parameters (inclination, mass-to-light ratio and two parameters
for the dark matter halo) that need to be determined, then $\Delta
\chi^2$ for a 68.3\% confidence interval is 4.7 \citep{press+86}, or
0.003 per data point for 1615 observables (960 $A_{lm}$ and 655
long-slit target observables). Therefore in the absence of systematic
errors, any models with a $\chi^2$ of more than 0.003 per data point
greater than the minimum $\chi^2$ per data point can be ruled out
because we know they must definitely be outside the 68.3\% confidence
range around the true best model. Table \ref{tab:alms+ls_chi2} shows
that the density observables most prefer an inclination of 75$\degree$
but the long-slit observables most prefer an inclination of
60$\degree$, in both potentials. Considering the data altogether, both
potentials prefer an inclination of 75$\degree$. Generally the
dynamical potential is preferred over the X-ray potential except for
in the most favoured inclination, where $VC1$ and $VC2$ are equally
preferred. As the remaining models achieve a $\chi^2$ of more than
0.003 per data point away from the minimum $\chi^2$ of 0.304 per data
point, we would rule them out with 68.3\% confidence using the $\Delta
\chi^2$ approach.

\begin{table}
  \centering
  \begin{tabular}{c c c c c c}
    \hline
    \hline
    Potential          &Slit PA ($\degree$)    &$v$     &$\sigma$  &$h_3$   &$h_4$ \\
    \hline
    X-ray               &105                    &0.131   &-0.196    &-0.013  &0.032\\
    ($VC1$)            &127                    &-0.005  &-0.048    &-0.036  &-0.179\\
                       &133                    &0.110  &-0.015    &-0.014  &0.165\\
                       &173                    &0.041   &-0.170    &0.033  &-0.003\\
                       &All                    &0.073   &-0.111    &-0.007  &0.009\\
    Dynamical          &105                    &0.041   &0.113     &-0.012   &0.149\\
    ($VC2$)            &127                    &-0.043  &0.086     &-0.013  &-0.141\\
                       &133                    &-0.032  &0.228     &0.018   &0.188\\
                       &173                    &0.038   &0.128     &0.031   &0.067\\
                       &All                    &0.019   &0.139     &0.006   &0.073\\
    \hline
  \end{tabular}
  \caption{Systematic differences between model and observed $v$,
    $\sigma$, $h_3$ and $h_4$, averaged over each of the first four
    slits for radii outside 4'', and then averaged over all the
    slits. \label{tab:sysdiff}}
\end{table}

We also quantify the systematic difference, $\Delta S$, between the
model and data in both potentials:
\begin{equation}\label{eq:sysdiff}
  \Delta S = \frac{1}{n}\sum_{j=1}^{j=n} \frac{K_{M,j} - K_{O,j}}{\epsilon_{K,j}}
\end{equation}
where $K$ represents the observable $v$, $\sigma$, $h_3$ or $h_4$, $M$
is the model value, $O$ is the observed value, $n$ is the number of
observables, and $\epsilon_{K,j}$ is the error on the observed
value. $\Delta S$ is calculated for each moment averaged over each of
the first four slits (the remaining three slits are only reflections
of the first three) for radii outside 4'', and then averaged over all
the slits. Only the results in the most favoured inclination of
75$\degree$ are shown in Table \ref{tab:sysdiff}. In $VC1$, the
magnitude of the model mean velocity is a little higher, while the
velocity dispersion is systematically a little lower than the
observations. The systematic differences in the $h_3$ and $h_4$
moments are smaller. In $VC2$, the systematic differences are biggest
also in the velocity dispersion, which is higher than the observations
on average, and in the $h_4$, which are also systematically
higher. Systematic differences between the model and the data seem
similar in both potentials and are also comparable to deviations used
to rule out models ($\sqrt{\Delta \chi^2 = 0.003} = 0.055$ per data
point). This implies that neither mass model is exactly correct and
that the true mass distribution lies somewhere in between, or that
systematic effects (e.g. triaxiality) play a role. Therefore the
$\Delta \chi^2$ approach should be used with caution.

\begin{figure}
\centering
\includegraphics[width=6.8cm,angle=-90]{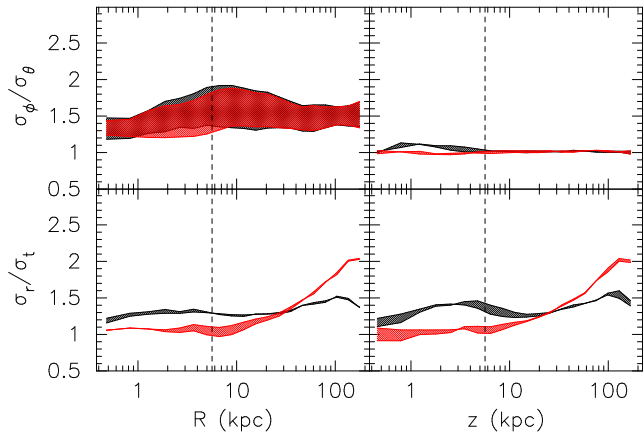}
\vspace{-1.0cm}
\caption{Intrinsic kinematics: The left set of panels show the ratio
  of the radial velocity dispersion to the tangential velocity
  dispersion (bottom), and the ratio of the azimuthal velocity
  dispersion to the meridional velocity dispersion (top), along
  $R$. The right set of panels show the same along $z$. The black
  region corresponds to the models carried out for various
  inclinations in $VC1$ and the red region corresponds to the models
  carried out for various inclinations in $VC2$, fitting photometric
  and long-slit constraints only. The dashed black line shows the
  radial extent of the kinematic
  constraints.\label{fig:alms+ls_intmoms}}
\end{figure}

\subsubsection{The orbital structure}\label{subsub:alms+ls_orb}

As we have assumed an oblate, axisymmetric system, we can average the
intrinsic kinematics over the azimuthal angle, $\phi$. This reduces
the spatial coordinates to $R$ in the equatorial plane, which is the
major axis of the system, and $z$ in the meridional plane, the minor
axis of the system.  Figure \ref{fig:alms+ls_intmoms} shows the ratio
of radial to tangential velocity dispersions ($\sigma_r/\sigma_t$),
and the ratio of azimuthal to meridional velocity dispersions
($\sigma_{\phi}/\sigma_{\theta}$) along $R$ and $z$.  The tangential
velocity dispersion is defined as $\sigma_t = \sqrt{(\sigma_{\theta}^2
  + \sigma_{\phi}^2)/2}$. Within the radial extent of the kinematic
constraints, there is a bias towards radial orbits in $VC1$ along $R$
and $z$ with $(\sigma_r/\sigma_t)_{max} \sim 1.5$. In the same region
the orbital structure in $VC2$ is almost isotropic along $R$ and
$z$. On average the ratio of the radial to tangential velocity
dispersions is 1.2--1.3 higher in $VC1$ than in $VC2$ throughout. The
two components of tangential velocity dispersions have similar
contributions in both potentials along $z$, and in the equatorial
plane the azimuthal dispersions dominate. In both potentials,
variation with inclination is greatest in the ratio between the
azimuthal and meridional dispersions along $R$.

\begin{table}
  \centering
  \begin{tabular}{c c c c c c c}
    \hline
    \hline
    $V_c$              &$\chi^2_{\textrm{alms}}$ &$\chi^2_{\textrm{long-slit}}$  &$\chi^2$        &$-LH$     &$-F$\\
    (1)                &(2)                    &(3)                          &(4)             &(5)       &(6)\\
    \hline
    $VC2$              &0.268                 &0.510                        &0.366           &3107.0   &4232.3\\
    $VC3$              &0.201                 &0.496                        &0.321           &3097.1   &4087.1\\
    $VC4$              &0.196                 &0.489                        &0.315           &3087.0   &3988.7\\
    $VC5$              &0.236                 &0.480                        &0.335           &3079.7   &4006.0\\
    $VC6$              &0.213                 &0.818                        &0.458           &3067.2   &4091.3\\
    \hline
  \end{tabular}
  \caption{Best-fit NMAGIC models for NGC 4649 for an inclination of
    75$\degree$ in a range of circular velocity curves: (1) Circular
    velocity curve, (2) $\chi^2$ per $A_{lm}$ target observable, (3)
    $\chi^2$ per long-slit target observable, (4) $\chi^2$ per target
    observable, (5) log-likelihood of PNe belonging to particle
    LOSVDs, and (6) the merit function. \label{tab:all_chi2}}
\end{table}

\begin{figure}
\centering
\includegraphics[width=5.0cm,angle=-90]{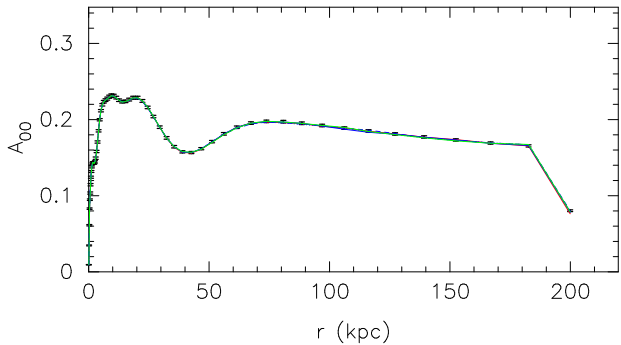}
\caption{Fit to the first moment of the $A_{lm}$s or differential
  stellar mass distribution for an inclination of 75$\degree$ in
  potentials $VC2$ (solid, red), $VC3$ (dotted, blue), $VC4$ (solid,
  blue), $VC5$ (dashed, blue), and $VC6$ (green). \label{fig:all_alms}}
\end{figure}

\subsection{Models fitting density, long-slit kinematic constraints
  and planetary nebula line-of-sight velocities}\label{sub:all}

Now we will carry out models including also the PN kinematics, so that
we can check whether the PN kinematics are consistent with the halo in
the mass determinations of \cite{das+10} and \cite{shen+10}. As both
potentials have almost the same halo we first carry out a model in the
potential of \cite{shen+10} ($VC2$). We find that the model velocity
dispersions are systematically too high and therefore repeat models in
a range of potentials with less massive haloes (see Figure
\ref{fig:vcirc} and Table \ref{tab:all_chi2}). Incorporating the PNe,
the models prefer a halo with a significantly lower circular velocity
of $\sim 463$ km/s compared to a value of 570 km/s at 45 kpc found in
\cite{shen+10}.

\begin{figure}
\centering
\includegraphics[width=6.5cm,angle=-90]{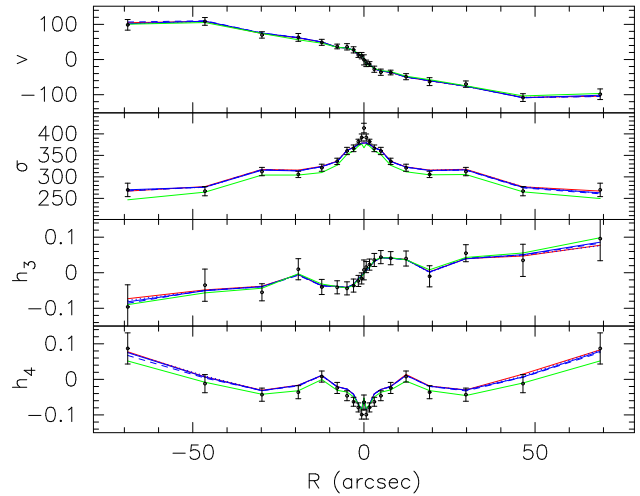}
\caption{Fits to the long-slit data: From the top down are shown mean
  velocity, velocity dispersion and the Gauss-Hermite moments $h_3$
  and $h_4$ for an inclination of 75$\degree$ and for the slit placed
  along the major axis. Black circles show the data and the lines show
  fits in potentials $VC2$ (solid, red), $VC3$ (dotted, blue), $VC4$
  (solid, blue), $VC5$ (dashed, blue), and $VC6$
  (green).\label{fig:all_ls}}
\end{figure}

\begin{figure*}
\centering
\includegraphics[width=11cm,angle=-90]{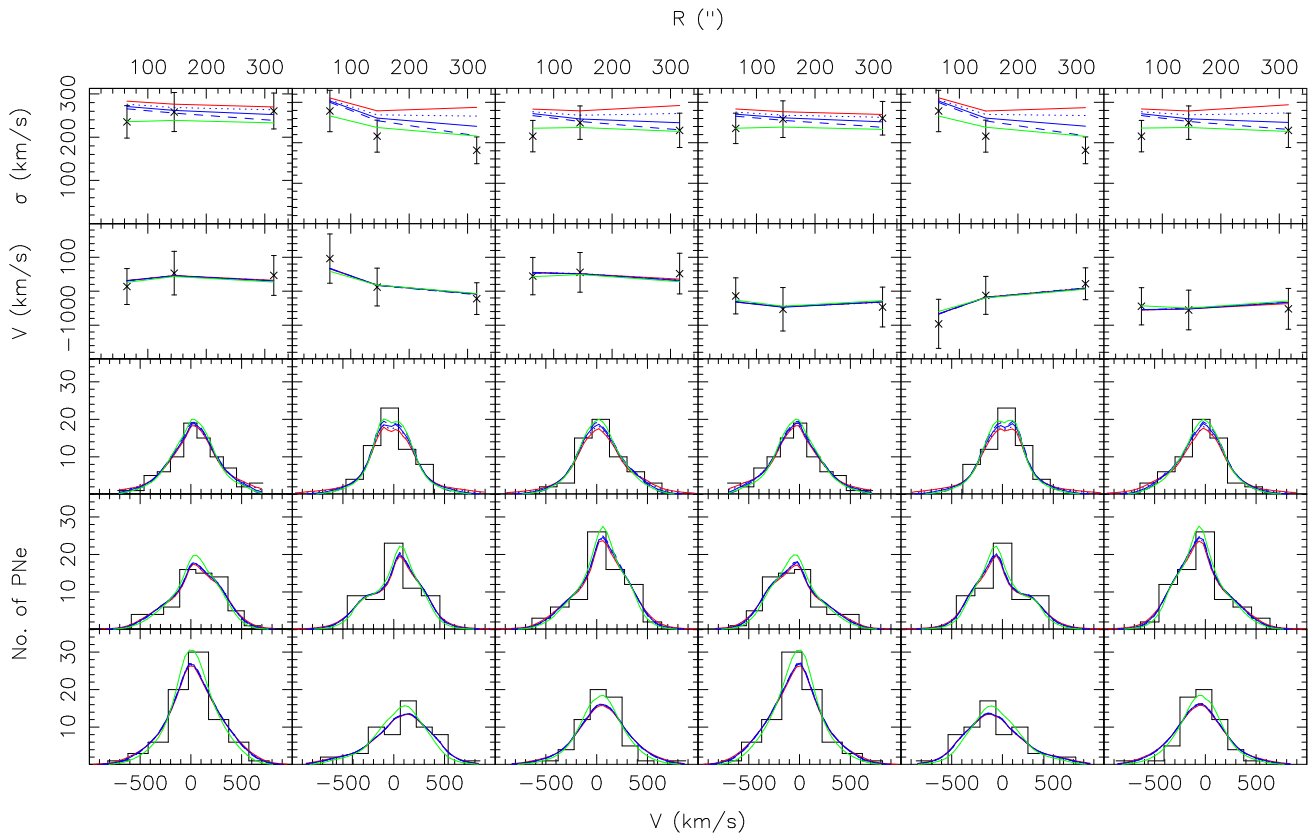}
\caption{Fits to the PN data for an inclination of 75$\degree$. Each
  of the plots in the bottom three rows shows the LOSVD in
  segments. Going upwards are segments at radii of 64'', 145'' and
  316'' and going towards the right are angular segments centred on
  0$\degree$ (major axis), 60$\degree$, 120$\degree$, 180$\degree$,
  240$\degree$, and 300$\degree$. The kinematics along the latter
  three segments are reflections of the kinematics along the first
  three segments because of the oblate, axisymmetry imposed on the
  observational constraints. The top two sets of panels show the mean
  velocity and velocity dispersion profiles along the angular
  segments. The fits are in potentials $VC2$ (solid, red), $VC3$
  (dotted, blue), $VC4$ (solid, blue), $VC5$ (dashed, blue), and $VC6$
  (green).\label{fig:all_pnelh}}
\end{figure*}

\subsubsection{Fits to the observables}\label{subsub:all_fitobs}

Figures \ref{fig:all_alms}, \ref{fig:all_ls} and \ref{fig:all_pnelh}
show the fits to the first moment of the $A_{lm}$s, the fits to the
Gauss-Hermite moments along the major axis and the fits to the PN LOS
velocities, in $VC2$, for an inclination of 75$\degree$. Table
\ref{tab:all_chi2} shows the statistics of the fits obtained for the
new model incorporating the PN data. Even though visually the fits
look very much the same in the $A_{lm}$s and long-slit, the $\chi^2$
values for the fit to the $A_{lm}$s and long-slit are slightly higher
than before, though still much less than 1. This shows that in this
potential, by trying to fit the PNe, the fits to the $A_{lm}$s and
long-slit are slightly compromised. Looking at the PN kinematics, it
appears that the mean velocity profiles are fit well but the velocity
dispersions of the model are systematically higher than that measured
by the PNe. This implies that the PNe are not consistent with the
outer part of $VC2$, and therefore also $VC1$ (both potentials agree
very well outside $\sim 12$ kpc).

\subsubsection{Fits in less massive haloes}\label{subsub:all_vclow}

As the PN dispersions are systematically lower than predicted by the
models, we also investigate additional circular velocity curves with
less massive haloes (Figure \ref{fig:vcirc}). We fit a straight line
to $VC2$ outside 7.6 kpc and obtain a slope of 0.212. We create
three additional curves where the outer slopes are 0.106 ($VC3$), 0.0
($VC4$) and -0.106 ($VC5$). Finally we consider a curve that follows
$VC1$ until 4.9 kpc and then also has an outer slope of 0.0
($VC6$). The model fits in these potentials are overplotted on Figures
\ref{fig:all_alms}, \ref{fig:all_ls}, \ref{fig:all_pnelh} and the
statistics of the fits are given in Table \ref{tab:all_chi2}. The
$\chi^2$ values show that the $A_{lm}$s and long-slit data most prefer
$VC4$, and the likelihood values show that the PNe most prefer
$VC6$. The merit function $F$ is a combination of the $\chi^2$ values,
likelihood and entropy, and this is a maximum in the potential
$VC4$. Therefore we consider this the best of the potentials
tried. The differences between the models in the $A_{lm}$s and fits to
the long-slit kinematics are small except in $\sigma$ and $h_4$, which
are systematically lower in $VC6$ compared to potentials
$VC2$--$VC5$. Figure \ref{fig:all_pnelh} shows that the PN velocity
dispersions are however very sensitive to the mass in the halo.

\begin{figure}
\centering
\includegraphics[width=6.8cm,angle=-90]{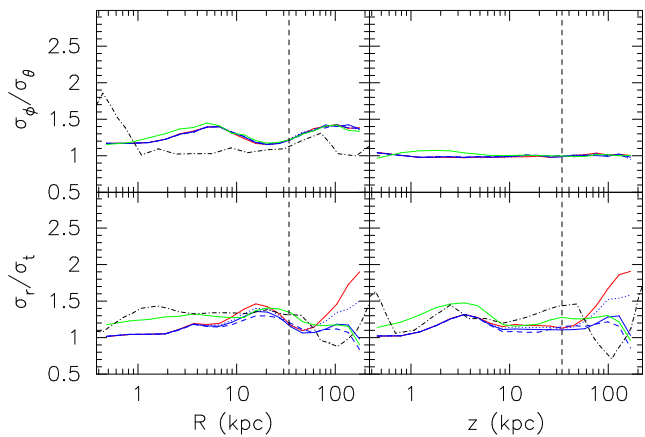}
\vspace{-1.0cm}
\caption{Intrinsic kinematics: The left set of panels show the ratio
  of the radial velocity dispersion to the tangential velocity
  dispersion (bottom), and the ratio of the azimuthal velocity
  dispersion to the meridional velocity dispersion (top), along
  $R$. The right set of panels show the same along $z$. The
  dash-dotted black line shows the results of \protect\cite{shen+10}
  who fit GC kinematics instead of PN kinematics. The fits are done in
  potentials $VC2$ (solid, red), $VC3$ (dotted, blue), $VC4$ (solid,
  blue), $VC5$ (dashed, blue), and $VC6$ (green). The dashed black
  line shows the radial extent of the kinematic
  constraints. \label{fig:all_intmoms}}
\end{figure}

\subsubsection{The orbital structure}\label{subsub:all_orb}

Figure \ref{fig:all_intmoms} shows that within $\sim 6$ kpc the
intrinsic velocity dispersions are isotropic to mildly radial along
$R$ and $z$ in potentials $VC2$--$VC5$ and moderately radial in $VC6$,
consistent with the findings in Section
\ref{subsub:alms+ls_orb}. Further out until the radial extent of the
PN kinematic constraints, the orbital structure in all potentials
becomes moderately radial along $R$ and more isotropic along $z$. The
two components of tangential velocity dispersions have similar
contributions in all the potentials, and in the equatorial plane
the azimuthal dispersions dominate over meridional velocity
dispersions.

\section{DISCUSSION}\label{sec:discuss}

Here we discuss what we have learned about the dark matter halo,
orbital structure and inclination of the stellar system in NGC
4649. We also discuss whether the PNe and GC systems are dynamically
consistent with each other, and whether the X-ray mass distributions
can be used to determine dark matter mass fractions and orbital
structures in massive elliptical galaxies.

\subsection{The dark matter halo of NGC 4649}\label{sub:disc_dm}

Figure \ref{fig:vcirc} shows the circular velocity curves that we have
probed in this work, compared to other recent determinations from
Chandra observations \citep{hump+06,hump+08} and a combination of
Chandra and XMM-Newton observations \citep{nag+09}. 

Our models fitting only density and long-slit kinematic constraints
cannot distinguish between the circular velocity curves of
\cite{das+10} and \cite{shen+10}, which differ only in the central
$\sim$ 12 kpc. A look at the systematic differences between the models
and observations in these potentials suggests that the true circular
velocity curve in the central $\sim 12$ kpc probably lies somewhere
between 425--500 km/s.

Models created incorporating additionally PN kinematics
\citep{teod+11} prefer a circular velocity curve that is flat outside
$\sim 12$ kpc at a value of $\sim 463$ km/s. This is most consistent
with the X-ray determination of \cite{nag+09}, slightly higher than
the determinations of \cite{hump+06,hump+08}, and lower than the
determinations of \cite{das+10} and \cite{shen+10}.  The sensitivity
of the PN velocity dispersions to the circular velocity curve in the
halo may be a result of a light density profile that falls off
approximately as -3 and an almost flat circular velocity
curve. \cite{gerhard93} showed that for such systems, the projected
velocity dispersions are constant and independent of anisotropy. The
constant circular velocity $V_c$ is then related to the constant
projected velocity dispersion $\sigma_P$ by $ V_c
=\sqrt{3}\times\sigma_P$. The average velocity dispersion of the PNe
outside $\sim 70$'' is $\sim 240$ km/s, therefore predicting a
circular velocity in the halo of 416 km/s. This lies in between the
value of 383 km/s of $VC6$, the halo most preferred by just the PNe,
and the value of 463 km/s of $VC4$, the halo most preferred
considering all the data together.

\begin{figure}
\centering
\includegraphics[width=1.0\linewidth]{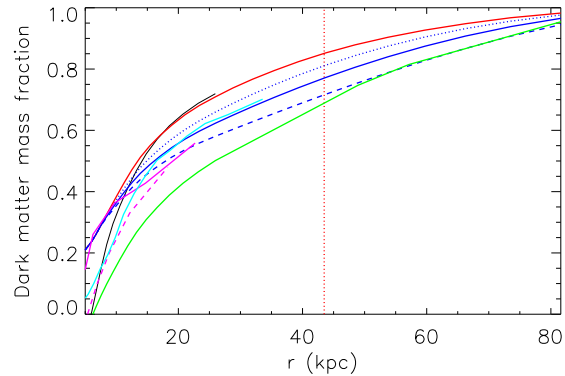}
\caption{Dark matter mass fraction of NGC 4649 according to potentials
  $VC1$ (black), $VC2$ (red), $VC3$ (solid, blue), $VC4$ (dotted,
  blue), $VC5$ (dashed, blue), and $VC6$ (green). For comparison, the
  dark matter mass fractions calculated from circular velocity curves
  obtained by \protect\cite{hump+06} (solid, pink),
  \protect\cite{hump+08} (dashed, pink), and \protect\cite{nag+09}
  (cyan) are also overplotted.\label{fig:dm}}
\end{figure}

To estimate the dark matter mass fraction, we first calculate the
luminosity enclosed within radius $r$ by integrating over the
luminosity density profile obtained from the spherical deprojection of
the surface-brightness profile. To obtain the mass in stars we
multiply this with a constant stellar population mass-to-light ratio
estimated in \cite{teod+11}. They used ages and metallicities in
\cite{trag+00} and the evolutionary population synthesis models of
\citet{maras98,maras05} for a Salpeter initial mass function with a
lower mass limit of $0.1M_{\odot}$. Their value of 10 in the $B$ band
corresponds to a value of about 7.6 in the $V$ band. Assuming that the
mass in gas is negligible, we subtract the mass in stars from the
total mass. Figure \ref{fig:dm} shows the dark matter mass fractions
corresponding to the potentials we have explored and the potentials
from the literature shown in Figure \ref{fig:vcirc}. The dark matter
mass fraction corresponding to our best potential $VC4$ is $\sim$ 0.39
at 10.5 kpc (1$R_e$), 0.5 at about 16 kpc ($1.5R_e$) and $\sim 0.78$
at the radius of the last PN ($\sim$45 kpc $\approx$ 4$R_e$). Figure
\ref{fig:dm} shows that our determination is most similar to that of
\cite{nag+09} as for the circular velocity curve.

There are also several determinations of dark matter fractions for
other elliptical galaxies in the literature. \cite{nag+09} analysed
Chandra and XMM-Newton observations for a sample of 22 elliptical
galaxies and found equality between dark matter and luminous matter at
around $3R_e$ and a dark matter mass fraction of around 0.66 at
$\sim6R_e$. \cite{gerhard+01} and \cite{thomas+07} found dark matter
mass fractions within $1R_e$ of 10--40\% and 10--50\% respectively,
for samples of nearby and Coma cluster elliptical
galaxies. \cite{gerhard+01} found equality between the mass
contributions of the dark matter and luminous components at 2--4$R_e$,
and for the Coma galaxies with sufficiently spatially extended data
(generally the less luminous ellipticals), \cite{thomas+07} found dark
matter mass fractions between 65--75\% at 4$R_e$.

\cite{treu+04} used a combined lensing and stellar dynamical approach
to obtain the dark matter mass fractions of elliptical and lenticular
galaxies up to a redshift of 1, and found projected values of
0.15--0.65 within a cylinder of radius 1$R_e$. For massive early-type
galaxies taken from the Sloan Lens ACS Survey (SLACS), \cite{auger+09}
and \cite{auger+10} found that the dark matter fraction within $R_e/2$
ranges between $\sim 0.3$--0.7 assuming a Chabrier stellar IMF (the
Salpeter stellar IMF gives lower dark matter mass fractions in their
study that are sometimes negative), with more massive galaxies having
higher dark matter mass fractions. \cite{grillo+10} used simple mass
models to estimate a dark matter mass fraction of $0.64_{0.11}^{0.08}$
projected inside a cylinder of radius 1$R_e$ for a sample of
approximately 170000 massive, elliptical galaxies observed in SDSS,
from which the SLACS sample is obtained. We estimate the effect of
projection on the dark matter mass fraction by assuming an NFW density
profile for the dark matter profile, with a virial radius of 300 kpc
and a concentration of 10 (typical for elliptical galaxies). The ratio
of the dark matter mass within a cylinder of length two virial radii
along the LOS and radius $R_e$, to that within a sphere of radius
$1R_e$ is about 2. Assuming that the ratio of the stellar mass between
the two regions is almost 1, then the dark matter mass fraction
calculated within a sphere of radius $1R_e$ would be lower, for
example 0.47 instead of 0.64.

\cite{onorbe+07} analysed the mass and velocity distributions of
elliptical-like objects at zero redshift in a set of self-consistent
hydrodynamical simulations set in the current cosmological
paradigm. They found that the objects are embedded in massive, dark
matter haloes with dark matter mass fractions ranging between 0.3--0.6
at 1$R_e$.

To summarise, there is a range in the dark matter mass fractions at
0.5--$1R_e$ in the literature, and the value we obtain for NGC 4649 is
near the middle of this range. Further out, the dark matter mass
fractions obtained by \cite{gerhard+01}, \cite{thomas+07} and
\cite{nag+09} suggest on average a more diffuse dark matter halo than
the one we find for NGC 4649. This may be because their samples
include elliptical galaxies at a range of luminosities, while massive
elliptical galaxies like NGC 4649 may have more massive dark matter
haloes \cite[e.g. ][]{cap+06,auger+10}.

\subsection{Orbital structure in NGC 4649}\label{sub:disc_orb}

The central $\sim 6$ kpc of NGC 4649 has an isotropic ($\beta = 1 -
\sigma_t^2/\sigma_r^2 \sim 0$) to mildly radial ($\beta \sim 0.4$)
orbital structure according to the dynamical potential of
\cite{shen+10} and the X-ray potential of \cite{das+10}, between which
we are unable to distinguish. If we assume that the true mass
distribution in the central $\sim 12$ kpc lies somewhere between these
two, as suggested by the systematic differences between the models and
the observations in the two potentials, then we can infer an orbital
structure in this region of $\beta \sim 0.2\pm 0.2$. Using
additionally the PN constraints and exploring several potentials, we
find that the orbital structure outside $\sim 4$ kpc becomes slightly
more radial ($\beta \sim 0.5$) along $R$, but more isotropic along
$z$, with little dependence on the exact halo assumed. Along $R$, the
azimuthal velocity dispersions are slightly higher than the meridional
velocity dispersions throughout, indicating that the stellar system
may be flattened by a meridional anisotropy in the velocity dispersion
tensor \citep{dehnen+93a,dehnen+93b,thomas+09b}.  \cite{thomas+09b}
also use axisymmetric toy models to show that flattening by meridional
anisotropy maximises the entropy for a given density
distribution. Along $z$, the azimuthal and meridional velocity
dispersions are equal, as one would expect for an oblate, axisymmetric
system.

There is no general consensus on the orbital structure in the outer
parts of elliptical galaxies. Dynamical models fitting outer
kinematics of intermediate-luminosity elliptical galaxies have found a
moderately radial orbital structure in the halo of NGC 4697
\citep{lorenzi+08}. In NGC 3379, \cite{lorenzi+09} found that systems
with both isotropic and moderately radial orbital structures were
consistent with the data, and \cite{nap+09} found moderately radial
anisotropy in the halo of NGC 4494.  Dynamical models fitting outer
kinematics in more massive elliptical galaxies such as NGC 1399
\citep[][the GCs used are believed to trace the stellar kinematics in
this galaxy]{schub+10} and NGC 4374 \citep{nap+10} have found mildly
radial and isotropic orbital structures in the halo respectively. For
the elliptical galaxies in the Coma cluster, which have a range of
luminosities, \cite{thomas+07} found mild radial anisotropy along the
major axis and sometimes tangential anisotropy along the minor axis.

From simulations, \cite{abadi+06} found that the outer haloes of
massive ellipticals are strongly radial due to smaller galaxies that
have been accreted on to the central object. The analysis of
\cite{onorbe+07} of elliptical-like objects at zero redshift also
found a radial orbital structure for the stars at an almost constant
value of $\beta \sim 0.5$ throughout. \cite{thomas+09b} analysed the
orbital structure of collisionless disc merger remnants from
\cite{naab+03} and found them to be strongly radially anisotropic.

To summarise, dynamical models show that the orbital structure in the
halo of massive elliptical galaxies is isotropic to quite radial, but
less so than expected from simulations. \cite{thomas+09b} arrived at
a similar conclusion and suggested that this could be due to gas
dissipational effects.

\subsection{The inclination of NGC 4649}\label{sub:disc_inc}

The surface-brightness and long-slit kinematic constraints in both
potentials $VC1$ and $VC2$ prefer an inclination of 75$\degree$ out of
the four inclinations we have probed. As we only explore a coarse grid
of inclinations at intervals of $15\degree$, we can attach an error of
$7.5\degree$ to our best inclination. Thus it appears that even if the
X-ray potential may not be completely correct throughout, it may be
used to find an approximate value for the inclination of the
system. This appears to be contrary to the work done for example by
\cite{kraj+05}, who found a degeneracy in the determination of the
inclination in their construction of axisymmetric models for the
elliptical galaxy NGC 2974. However, we have assumed the same
spherical total potential in all the inclinations for our first set of
models, and only the stellar distribution was axisymmetric and varied
according to the inclination. To understand this issue in more depth,
a range of mass profiles need to be explored in each inclination to
see whether equally good but different mass profiles can be found for
each of the inclinations.

\begin{figure*}
\centering
\subfloat[]{
  \includegraphics[width=0.5\linewidth]{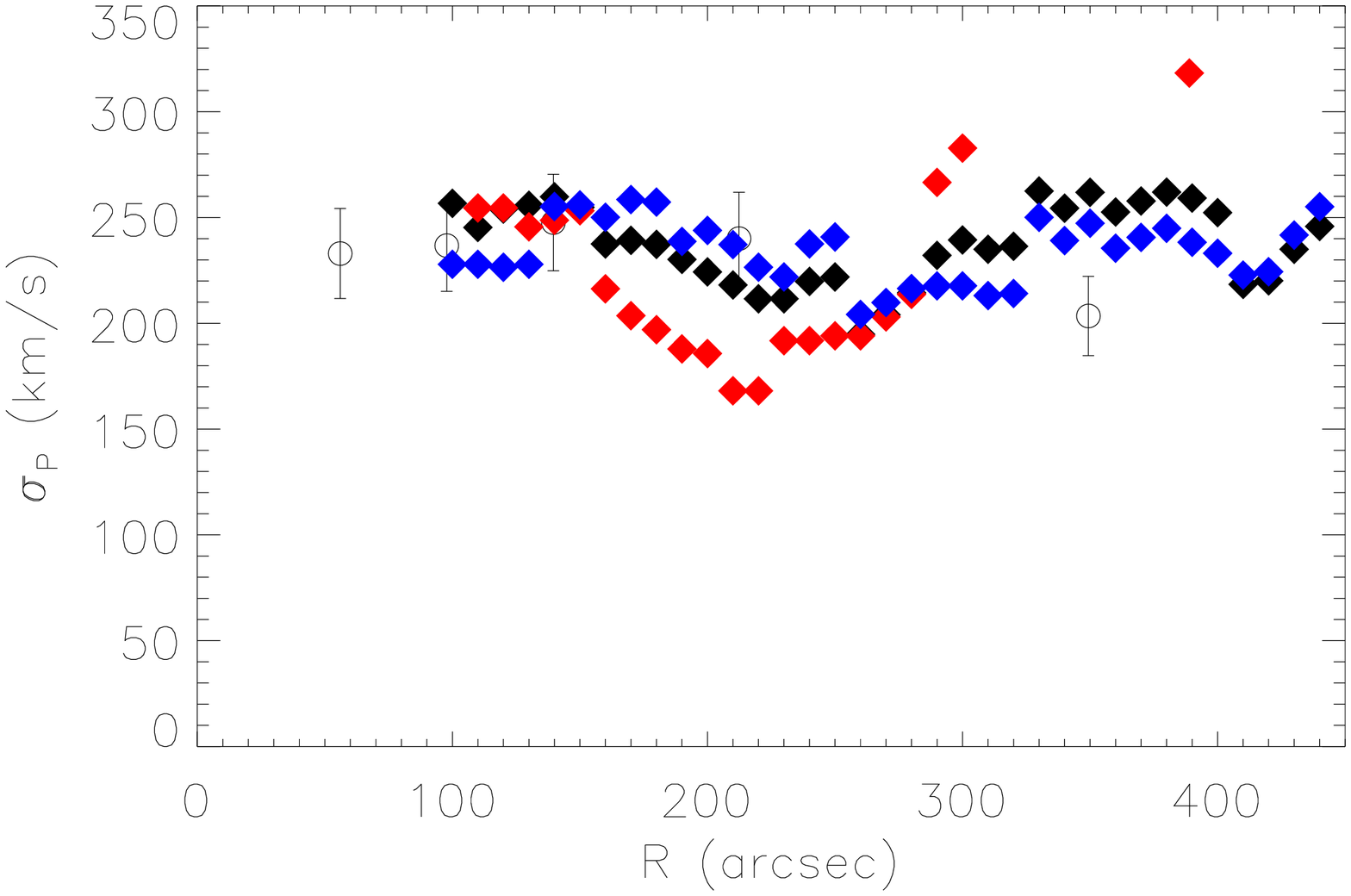}
}
\subfloat[]{
  \includegraphics[width=0.5\linewidth]{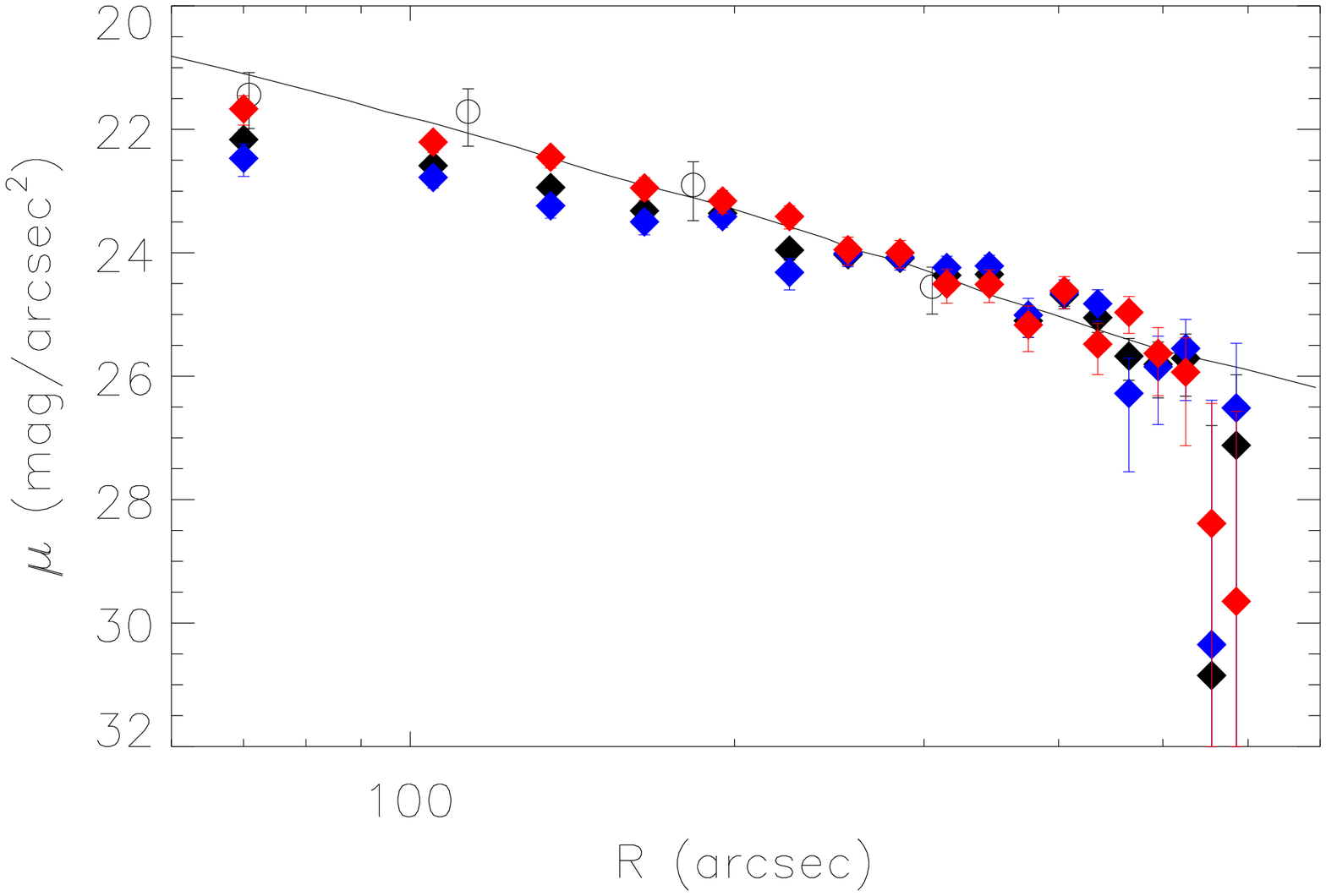}
}
\caption{Properties of the GC populations: (a) Velocity dispersion
  profiles of all GCs (filled, black diamonds), red GCs (filled, red
  diamonds) and blue GCs (filled, blue diamonds) compared to that
  measured by the PNe (open, black circles). The errors on the GC
  velocity dispersions are approximately twice as big as the errors
  shown on the PN velocity dispersions. (b) Surface number density
  profiles of all GCs (filled, black diamonds), red GCs (filled, red
  diamonds) and blue GCs (filled, blue diamonds) compared to the
  stellar surface-brightness distribution (solid, black line) and the
  surface number density of PNe (open, black circles). The GC points
  are not independent as they are obtained with a moving window
  average.\label{fig:gcscomp}}
\end{figure*}

\subsection{Are the PNe and GCs dynamically consistent with each other?}\label{sub:disc_pne}

The NMAGIC models show that the PN kinematics prefer a less massive
halo and a less radial orbital structure than that derived from models
using the stellar density profile and GC kinematics in
\cite{shen+10}. However if both the PN and GC systems are in
equilibrium, the same halo should be preferred although the orbital
structure may be different. To obtain a deeper insight into this
discrepancy, we examine the photometric and kinematic constraints
assumed for the PNe and GCs in more detail. Figure
\ref{fig:gcscomp}(a) compares the velocity dispersion profile
calculated in circular rings from the PNe to those for red, blue, and
all GCs calculated by \cite{hwang+08}. The errors bars on the GC
kinematics are not shown because they mask all the other points but
are approximately twice as big as the errors shown on the PN velocity
dispersions. One can see that between 100'' and 200'', the PN velocity
dispersions appear to be in agreement with the dispersions measured by
the blue GCs (and therefore the total sample because the blue GCs
dominate in numbers). The velocity dispersion of the red GCs decreases
to almost 150 km/s at $\sim$210'' and then increases dramatically to
about 320 km/s at $\sim$390''. The last PN velocity dispersion point
at $\sim$350'' calculated in a ring extending from 260--440'' has a
value of $203\pm 19$ km/s. Averaging all the GCs velocity dispersions
in this region gives $\sim 240 \pm 22$ km/s for 61 GCs, which would
correspond to the values fit by \cite{shen+10}. This is only just
consistent and therefore it is possible that the PNe and GCs trace
different kinematics.

Figure \ref{fig:gcscomp}(b) compares the surface-brightness
distribution of the stars with scaled surface number densities
calculated from PNe in \cite{teod+11} and GCs from
\cite{hwang+08}. Between 200--450'', the surface densities of the
whole GC population follows the stellar surface-brightness
distribution well. Further in, the GC density profile is shallower
than that of the stars and outside this region the density profile
falls off more steeply than the stars, though the associated errors
are larger. Therefore the density profile of the GCs appears to be
different from that of the stars.

The GCs could still be in equilibrium however but just form a separate
dynamical system in the same potential. This can be quantified using
the spherical second-order Jeans equation, relating the second-order
moments of the intrinsic velocity distribution to the density profile
of the stars, and the potential in which they move \citep{binney+87}:
\begin{equation}\label{eq:jeans}
  \frac{\diff}{\diff r}(\nu(r)\sigma_r^2(r)) +
\frac{2\beta(r)}{r}\nu(r)\sigma_r^2(r) +\nu (r)\frac{V_c^2}{r} = 0
\end{equation}
where $V_c$ is the circular velocity curve, $\sigma_r$ is the
intrinsic velocity dispersion of the tracer in the radial direction,
$\nu$ is the number density of the tracer, and the anisotropy $\beta =
1 - \sigma_t^2/\sigma_r^2$.

This equation shows that as the PNe and GCs are residing in the same
halo but have different density profiles and probably different
kinematics, for the GCs to be in equilibrium, they must also have a
different orbital structure. We can estimate the orbital structure
most easily for the case of a power-law density profile, constant
anisotropy and constant circular velocity curve. Therefore we fit a
power law to the surface-brightness measured by the stars outside
200'' and find a best-fit index of $n_{\textrm{stars}} = -2.2$. For
the globular clusters we find $n_{\textrm{gcs}} = -3.0$ (ignoring the
second last point, which seems unphysically low).  The power-law
indices of the intrinsic density profiles are then -3.2 and -4.0
respectively. The projected velocity dispersion is $\sim 240$ km/s
outside 200'', and we assume a circular velocity of 463 km/s in the
halo from our best potential $VC4$. Solving the Jeans equation and the
equation relating projected velocity dispersions to intrinsic velocity
dispersions, we find $\beta \sim -1$ for the GCs.  This is different
from the average value of $\beta \sim 0.4$ outside 200'' obtained by
\cite{shen+10}, who assume the stellar density profile for the
GCs. Our value agrees better with the modestly tangentially biased
velocity ellipsoid inferred by \cite{hwang+08} from spherical Jeans
equations using the true GC density profile and the mass distribution
from X-rays in \cite{hump+06}, which is more consistent with the mass
distribution we find from the PNe.

To summarise, we show that the GCs may be in equilibrium in the
potential of NGC 4649, but this equilibrium is dynamically distinct
from the stars and PNe. Therefore to infer the mass of the dark matter
halo, the correct density profile and kinematics need to be used.

\subsection{Are the mass distributions from X-rays accurate enough to determine dark matter mass fractions and orbital structures?}\label{sub:disc_xrays}

Using the photometric and long-slit kinematic constraints in the
dynamical potential from \cite{shen+10} and the X-ray potential from
\cite{das+10}, we find that the same inclination of 75$\degree$ is
preferred for the stellar system. In this inclination we are unable to
distinguish between them. As the X-ray potentials from
\cite{hump+06,hump+08} and \cite{nag+09} are similarly low in the
central $\sim 12$ kpc, we would expect a similar result if one of
these were used instead. Looking at the systematic differences between
the models and observations in the two potentials suggests that the
true mass distribution in the central $\sim 12$ kpc lies somewhere in
between. By assuming the X-ray mass distribution, a maximum systematic
error of 0.4 in $\beta$ is made, i.e. one finds a more radial orbital
structure.

From our results, we do not see compelling evidence for non-thermal
pressure contributions in the gas or multi-phase components in the gas
in the central $\sim 12$ kpc. If they exist, their effects are smaller
than would be inferred from the potential of \cite{shen+10}. This is
consistent with the work of \cite{brig+09}. They model the hot gas in
NGC 4649 using 2-D gas-dynamical computations and find that a
turbulent pressure is required in NGC 4649, but it has a much smaller
contribution than the thermal pressure.

In the halo, the mass distribution most preferred by incorporating the
PN data is less massive than that found in the X-ray analysis of
\cite{das+10}, slightly more massive compared to the halo found in
\cite{hump+06,hump+08}, and similar to that found by
\cite{nag+09}. The dark matter mass fractions derived in the halo
depend highly on the mass distribution assumed. The orbital structure
in the halo however does not change much between the potentials
explored. Still the important question arises: why do the X-ray mass
distributions differ in the outer regions? Looking at the temperature
and pressure profiles measured from Chandra and XMM-Newton
observations in \cite{das+10}, we find that at $\sim 25$ kpc, the
temperature profiles are consistent but the pressure profiles are
not. The pressure calculated from the Chandra data is higher than that
calculated from the XMM-Newton data, resulting in a flatter pressure
profile in the outer parts. Looking at Equation \eqref{eq:he} shows
that this will result in a lower circular velocity
curve. \cite{das+10} use both sets of observations but omit the final
points due to uncertainties associated with the deprojection, and
therefore do not use the Chandra point at 25 kpc. \cite{hump+06} do
use this point however. Therefore it seems that the outer slope of the
mass distribution from X-rays can be quite uncertain and possible
effects that need to be explored in more detail are deprojection
issues, metallicity gradients in the hot gas, and outflows.

To summarise, we believe that using the X-ray mass distribution may
lead to a systematically more radial orbital structure in the central
region. Models with the X-ray mass profile however are able to derive
the inclination of the stellar system and the orbital structure in the
halo. Therefore until the uncertainties in the derivation of X-ray
mass distributions are better understood, it is best to use them in
conjunction with a dynamical mass analysis.

\section{CONCLUSIONS}\label{sec:conc}

We have created dynamical models of the Virgo elliptical galaxy NGC
4649, using the highly flexible made-to-measure N-body code, NMAGIC,
and observational constraints given by surface-brightness data,
long-slit kinematics, and planetary nebula (PN) velocities. We explore
a range of potentials based on X-ray mass distributions in the
literature, which are similar in the central regions, but have
different outer slopes, and a dynamical potential derived from
globular cluster (GC) velocities and a stellar density profile. The GC
dynamical model prefers more mass in the central region compared to
the X-ray potentials, and is on the top end of the range of X-ray mass
profiles further out.

Our models are not able to differentiate between the X-ray and GC mass
profiles in the central $\sim 12$ kpc, and the systematic differences
suggest that the true circular velocity curve in the central $\sim 12$
kpc may lie somewhere between 425--500 km/s. Therefore if non-thermal
pressures or multi-temperature components exist in the central region,
their contribution is less than previously inferred.

Outside $\sim 12$ kpc, the observational constraints prefer a circular
velocity curve that is flat with a value of $\sim 463$ km/s, most
consistent with the X-ray determination of \cite{nag+09}. The PN
velocity dispersions are very sensitive to the circular velocity curve
in the halo, possibly as a result of a stellar density profile that
falls off approximately as -3 and an almost flat circular velocity
curve \citep{gerhard93}. The discrepancy between the halo mass
preferred by the PN kinematics and that corresponding to the GC
dynamical model shows that if the GCs are in equilibrium, they are
dynamically distinct from the stars and PNe. Therefore the correct
density profile and kinematics need to be used to infer the mass of
the dark matter halo.

We find a dark matter mass fraction of 0.39 at $1R_e$ for NGC 4649,
which is generally in agreement with the values in the literature. At
$4R_e$ we obtain a dark matter mass fraction of 0.78, suggesting a
more massive halo than typical for the samples analysed by
\cite{gerhard+01}, \cite{thomas+07} and \cite{nag+09}. This may be
because their samples include elliptical galaxies at a range of
luminosities, while massive elliptical galaxies like NGC 4649 may have
more massive dark matter haloes \cite[e.g. ][]{cap+06,auger+10}.

We find an orbital structure that is isotropic to mildly radial in the
central $\sim 6$ kpc, depending on the potential assumed. Further
out, we find that the orbital structure becomes slightly more radial
along $R$, but more isotropic along $z$, with little dependence on the
exact halo assumed. Along $R$, the azimuthal velocity dispersions are
slightly higher than the meridional velocity dispersions throughout,
indicating that the stellar system may be flattened by a meridional
anisotropy \citep{dehnen+93a,thomas+09b}. The orbital structure in the
halo of NGC 4649 is less radial than expected from simulations,
possibly due to gas dissipational effects \citep{thomas+09b}.

Assuming a mass distribution from X-rays leads to a systematically
more radial orbital structure in the central region, but recovers the
orbital structure in the halo. The inclination of the stellar system
is also recovered. It is apparent that until the uncertainties in the
derivation of X-ray mass distributions are better understood, they
should only be used alongside a more detailed dynamical mass analysis.

\section*{ACKNOWLEDGEMENTS}

PD was supported by the DFG Cluster of Excellence ``Origin and
Structure of the Universe''. RHM and AMT acknowledge support by the
National Science Foundation (USA) under grant 0807522. We would like
to thank H. Hwang for providing the surface number density and
velocity dispersion profiles of globular clusters belonging to NGC
4649.

\bibliographystyle{mn2e}
\bibliography{paper}

\appendix

\label{lastpage}

\end{document}